\documentclass[fleqn,usenatbib]{mnras}

\usepackage[T1]{fontenc}

\DeclareRobustCommand{\VAN}[3]{#2}
\let\VANthebibliography\thebibliography
\def\thebibliography{\DeclareRobustCommand{\VAN}[3]{##3}\VANthebibliography}

\usepackage{graphicx}	% Including figure files
\usepackage{amsmath}	% Advanced maths commands
\usepackage{amssymb}	% Extra maths symbols
\usepackage{longtable}

\usepackage{newtxtext,newtxmath} %Шрифты, похожие на Times New Roman (перенесено с места выше)

\newcommand{\di}{\mathrm{d}}

\newcommand{\Ep}{E_{\mathrm{p}}}
\newcommand{\Epi}{E_{\mathrm{p},i}}

\newcommand{\Eiso}{E_{\mathrm{iso}}}
\newcommand{\Sbolo}{S_{\mathrm{bolo}}}
\newcommand{\Sobs}{S_{\mathrm{obs}}}
\newcommand{\Sbolocor}{S_{\mathrm{bolo}}^{\mathrm{cor}}}
\newcommand{\kmsbyMpc}{\frac{\text{km/s}}{\text{Mpc}}}
\newcommand{\Omm}{\Omega_{\mathrm{m}}}
\newcommand{\OmL}{\Omega_{\Lambda}}
\newcommand{\Omk}{\Omega_{\mathrm{k}}}
\newcommand{\Deltal}{\Delta_{\mathrm{l}}}
\newcommand{\Deltau}{\Delta_{\mathrm{u}}}
\newcommand{\muagrb}{\mu_{\mathrm{GRB}}^{\mathrm{A}}}

\title[Solving the ICCP of LGRBs]{Solving the inverse cosmological calibration problem of gamma-ray bursts}

\author[S. I. Shirokov et al.]{
S. I. Shirokov,$^{1}$\thanks{E-mail: arhath.sis@yandex.ru}
R. I. Gainutdinov,$^{1,2}$\thanks{E-mail: roustique.g@gmail.com} 
N. Yu. Lovyagin,$^{2}$\thanks{E-mail: n.lovyagin@spbu.ru}
and V. L. Gorokhov$^{3}$
\\
$^{1}$SPb Branch of Special Astrophysical Observatory of Russian Academy of Sciences, 65 Pulkovskoye Shosse, St Petersburg 196140, Russia\\
$^{2}$Saint Petersburg State University, 7/9 Universitetskaya Nab., St Petersburg 199034, Russia\\
$^{3}$Saint Petersburg Electrotechnical Univeristy,
Ulitsa Professora Popova 5, 197376 St. Petersburg, Russia
}

\date{Accepted XXX. Received YYY; in original form ZZZ}

\pubyear{2022}

\begin{document}
\label{firstpage}
\pagerange{\pageref{firstpage}--\pageref{lastpage}}
\maketitle

\begin{abstract}
We have received a new physical characteristics fitting based on actual observational data from the Swift mission's long-duration gamma-ray bursts (LGRBs). We considered such characteristics as the Amati parameters for linear correlation ($\Eiso-\Epi$) and the $k$-correction for gravitational lensing and Malmquist bias (GLMB) effect. We used the Pantheon SN Ia catalogue and the standard $\Lambda$CDM model with a fixed Hubble constant of $H_0=70$ km/s/Mpc as the baseline for the Hubble function $\mu(z)$. 
In our paper, we formulated the inverse cosmological calibration problem (ICCP) in the non-parametric statistics framework. The ICCP involves fitting non-observable physical characteristics while assuming a fixed cosmological model. To solve this problem, we developed a new method that is resistant to non-Gaussian processes. This method is based on error propagation through the Monte Carlo method and the Theil-Sen method for linear regression estimate. We have demonstrated the stability and robustness of this assessment method.
The parameter estimates are as follows: $a=0.92^{+0.12}_{-0.12}$, $b=50.32^{+0.33}_{-0.32}$ without considering the GLMB effect, and $a=0.63^{+0.13}_{-0.14}$, $b=50.12^{+0.33}_{-0.31}$, and $k=1.98^{+0.25}_{-0.24}$ with the effect included. 
The proposed method can be applied to any other calibration sample of known standard candles, a calibrated sample of LGRBs, and the Hubble function $\mu(z)$. 
In the future, the ICCP idea can be used as an alternative cosmological test for estimating cosmological parameters, including the GLMB effect, or even for the selection of models, providing new information about the Universe. This can be done by analysing the residual values of observational data within the Bayesian statistics paradigm.
% in frameworks
\end{abstract}

% Anyone with this link can edit this project
% https://www.overleaf.com/4877733698fwphmnnxqktj
% Anyone with this link can view this project
% https://www.overleaf.com/read/bnptxcxvvjkf


%%%%%%%%%%%%%%%%%%%%%%%%%%%%%%%%%%%%%%%%%%%%%%%%%%

%%%%%%%%%%%%%%%%% BODY OF PAPER %%%%%%%%%%%%%%%%%%

% new citations:
%P. T. Boggs and J. E. Rogers, “Orthogonal Distance Regression,” in “Statistical analysis of measurement error models and applications: proceedings of the AMS-IMS-SIAM joint summer research conference held June 10-16, 1989,” Contemporary Mathematics, vol. 112, pg. 186, 1990.

% Обратная задача, случай варьирования a, b, k: 4.95657
% Обратная задача, случай варьирования a, b с фиксированным k=0: 5.29550
% Для прямой задачи: 5.43035

\section{Introduction}

Long gamma-ray bursts (LGRB) are large-scale explosive events that occur in distant galaxies and are observed in the highest energy part of the electromagnetic spectrum. The LGRB sources are related to explosions of massive core-collapse SN in distant galaxies~\citep{cano2017observer}, though up to now there is no satisfactory theory of the LGRB radiation origins~\citep{willingale2017gamma}.
They have a typical duration of a few seconds, although they can also be as long as milliseconds up to an hour. The idea of using LGRBs as standard candles (SCs) in cosmology is actively discussed in the last two decades~\citep{amati2002intrinsic,amati2008measuring,amati2019addressing,demianski2017cosmology,demianski2017cosmology2,lusso2019tension,yonetoku2004gamma,wang2011updated,wei2017gamma}.
Since LGRBs are extremely distant observable objects, using them as standard candles makes it possible to significantly extend the scale of cosmological distances. This makes it possible to use LGRBs to determine the parameters of cosmological models. One of the main cosmological tests is the distance--redshift diagram, also known as the Hubble diagram (HD). Today, the HD remains a relevant cosmological test, with which one can test various hypotheses or estimate cosmological parameters. Based on cosmological models, one usually describes theoretical the distance--redshift relation as a parametric function $d(z,\mathbf p)$, where $\mathbf p$ is a parameter vector. Thus, cosmology-independent determination of distances to objects with known redshift gives one a unique opportunity to verify, compare and probe parameters of cosmological models~\citep{baryshev2012fundamental}. 

The HD is a well-known and widely-used practical cosmological test~\citep{sandage1995astronimical,baryshev2012fundamental,shirokov2020high,shirokov2020b}. 
The first use of it can be considered as the discovery of the expansion of the Universe in 1929, described in the classic work of Edwin Hubble~\citep{Hubble1929}. The linearity of this Hubble law on the scales 1--300 Mpc was confirmed by Sandage at the Hale Telescope~\citep*[current results are presented in][]{Sandage2010}. At the end of the 20th century, the HD constructed for type Ia supernovae made it possible to discover the accelerated expansion of the Universe within the framework of the Friedmann--Lemaitre--Robertson--Walker metric. This led to the introduction of dark energy into the standard cosmological model (SCM)~\citep{Riess1998,Perlmutter1999}. However, there is still a wide discussion on cosmological models and their parameter values~\citep{aghanim2020planck,riess2018new,riess2020expansion,yershov2020distant}. 

Supernovae of the Ia type are classical observable standard candles, however, with current observable facilities, the limit redshifts for them of the order of $z \sim 2\div 3$. While LGRBs were watching up to $z \sim 10$~\citep{Amati2018}. Therefore, the inclusion of LGRBs in the scale of cosmological distances allows one to construct a HD that includes a deeper region of the Universe. Also, one can say that LGRBs are between supernovae and the cosmic microwave background radiation (CMBR), which is associated with probing cosmological parameters of very far ($z\sim 1\,000$), or very early Universe. 
% one can also mention quasars as far objects (not SC) here
Thus, the LGRB HD could be used as cosmological test that is a kind of ``connecting link'' between the SN and CMBR cosmology~\citep{Shirokov2020c}. 

We suppose that LGRB HD can be used as probing cosmological parameters $\mathbf p$ and comparing cosmological models, as well as the SNe HD. However, because of large variance and uncertainties of LGRB parameters, the LGRBs can be use as SCs in a mean sense only. The mean sense implies that we must use averaging characteristics (e.g., quantiles, or function fittings) for determining cosmological parameters.
In order for LGRBs to be considered as SC, a correlation between the energy of the gamma fluence and the distance to LGRBs is required. One such correlation is the linear dependence of the isotropic equivalent radiated energy $\Eiso$ on the peak energy of the spectrum $\Epi$~\citep{Amati2008}, also called the Amati relation.
The relation give one an opportunity to cosmology-independent measuring the distance by LGRBs with known redshift. However, it depends on two unknown parameters that have to be calibrated observationally. For calibration of the correlation, the LGRBs with known distances (and redshifts) are needed so that the first step of our study is in determining distances $d$ for subsample of LGRBs in near galaxies by cosmology-independent methods.

The Amati relation is a rather rough correlation, since, for example, it does not take into account the angle of the jet cone, under which we see this burst. A more accurate correlation that takes this angle into account is called the Ghirlanda relation~\citep{ghirlanda2004collimation,ghirlanda2007confirming}.
The main problem in using LGRBs for cosmological tests is getting rid of the dependence of the distances measured from the observed parameters on the choice of the cosmological model~\citep{Kodama2008}. Due to this dependence, the measured distances to the LGRBs require additional calibration.

In our article~\citet{Lovyagin2022}, we have showed a method for calibrating LGRBs with $z < 1.4$ for type Ia supernovae from the Pantheon catalogue~\citep{scolnic2018complete}. The method is consisted in the fact that the distance modulus $\mu $ -- redshift $z$ dependence for supernovae was approximated, and then used to calculate the LGRB distance moduli in the $\log \Eiso$ -- $\log \Epi$ plane. Next, using the estimated Amati parameters, we have plotted a HD for our LGRB sample from the Swift catalogue~\citep{gehrels2004swift}.
In this paper, the LGRB calibration by supernovae is carried out in a different way. 
We stick to the Bayesian statistics as a way to getting new information about Universe. 

Solving inverse cosmological calibration problem (ICCP) in the frameworks of fixed cosmological model, we can get fitting physical characteristics based on observational data. In case of LGRB these characteristics are the Amati parameters of linear correlation $\Eiso-\Epi$~\citep{amati2008measuring} and the $k$-correction for gravitational lensing and Malmquist bias (GLMB) effect~\citep{shirokov2020high}. We try to estimate these characteristics through fitting the Swift LGRB data by using the SCM as a basis. 

At the first stage, the cosmological parameters of the $\Lambda$CDM model are estimated from the HD for supernovae.
% and match the Planck results~\citep{aghanim2020planck}. 
The cosmological model with these parameters is then considered as fixed. 
Next, we estimate the parameters $a, b$ of the linear Amati relation and the parameter $k$. Thus, we find the optimal values of the parameters $a, b, k$, i.e. such that the LGRB distance moduli, calculated by using these parameters, are corresponded to the $\Lambda$CDM model with parameters, estimated from type Ia supernovae from the Pantheon catalogue. 

\section{Standard Cosmological Model}
\label{sec:cosm}

At the moment, the $\Lambda$CDM model with parameters $H_0=70,\,\OmL=0.7$ is accepted as the standard cosmological model (SCM), which is a special case of the Friedmann model, a non-stationary model of the Universe that satisfies the Einstein equations, which in a simplified form can be written as follows:
\begin{equation}
    G_{\mu\nu} + \Lambda g_{\mu\nu} = \kappa T_{\mu\nu},
    \label{eq:Einstein}
\end{equation}
where $G_{\mu\nu}$ is the Einstein tensor, $\Lambda$ is the cosmological constant, $g_{\mu\nu}$ is the metric tensor, $T_{\mu\nu}$ is the energy-momentum tensor, and $\kappa = 8\pi G / c^4$ is a constant. In the $\Lambda$CDM model, it is assumed that the Universe, in addition to baryonic matter, is also filled with cold dark matter (CDM) inaccessible to direct observation, as well as dark energy, which is related to the cosmological constant $\Lambda$ from the Einstein equations~\ref{eq:Einstein}.
The content of matter (visible and dark combined) is described by such a parameter as the density divided by the critical density:
\begin{equation}
    \Omm = \frac{\rho_{\mathrm{m}}}{\rho_{\mathrm{cr}}}, \quad \rho_{\mathrm{cr}} = \frac{3H_0^2}{8\pi G}
\end{equation}
For dark energy, the relative density is defined as follows:
\begin{equation}
    \OmL = \frac{\rho_{\Lambda}}{\rho_{\mathrm{cr}}} = \frac{\Lambda}{3} \Big( \frac{c}{H_0} \Big)^2.
\end{equation}
In addition to these two parameters, the curvature contribution is also introduced\footnote{In most papers, the quantity $\Omk$ is defined with the opposite sign. We, following~\citet{baryshev2012fundamental}, use just such a notation in which the density of curvature and the curvature itself have the same sign.} $\Omk$:
\begin{equation}
    \Omm + \OmL - \Omk = 1.
    \label{eq:friedmann}
\end{equation}
Space-time curvature occurs if the average density of the universe differs from the critical one:
\begin{itemize}
    \item $\Omk = 0 \;\Leftrightarrow \overline{\rho} = \rho_{\mathrm{cr}}$ -- flat, open space\-/time;
    \item $\Omk < 0 \;\Leftrightarrow \overline{\rho} < \rho_{\mathrm{cr}}$ -- negative curvature, open space\-/time;
    \item $\Omk > 0 \;\Leftrightarrow \overline{\rho} > \rho_{\mathrm{cr}}$ -- positive curvature, closed space\-/time.
\end{itemize}
Thus, there are 4 quantities to parameterise the $\Lambda$CDM model:
\begin{itemize}
    \item $H_0$ -- Hubble constant, a linear coefficient in Hubble's law that relates the distance to an extragalactic object and the speed of its receding (or approaching). Standard units: $\kmsbyMpc$;
    \item $\Omm$ -- density of matter (visible together with dark matter) reduced to critical density. Dimensionless value;
    \item $\OmL$ -- density of dark energy reduced to critical density. Dimensionless value;
    \item $\Omk$ -- the contribution of curvature to the total average density, $\Omk = \Omm + \OmL - 1$ (in most papers it is defined with the opposite sign). Dimensionless value.
\end{itemize}

Of these parameters, three (relative densities) are related by Equation~(\ref{eq:friedmann}), so this model actually has three degrees of freedom, not four. But usually, for the sake of clarity, the above four quantities are written out in their entirety. In this case, in the SCM framework $\Omk=0$, and thus, there are two degrees of freedom.
These parameters are included in the equations (\ref{eq:metricdist}, \ref{eq:h}, \ref{eq:del}, \ref{eq:mu}) relating redshift to distance. Accordingly, in order to test any cosmological model (in particular, to estimate the parameters of the $\Lambda$CDM model), we need to build a cosmological test, e.g., a HD (the distance -- redshift dependence) obtained from observations. Expression for metric distance:
\begin{equation}
    l(z) = \frac{c}{H_0}\frac{1}{\sqrt{|\Omk|}}\;\mathrm{sn}_k \bigg( \sqrt{|\Omk|} \int\limits_0^z \frac{\di z'}{h(z')} \bigg)
    \label{eq:metricdist}
\end{equation}
where $\mathrm{sn}_k(x) = \mathrm{sin}(x)$, if $\Omk > 0$; $\mathrm{sn}_k(x) = x$, if $\Omk = 0$ and $\mathrm{sn}_k(x) = \mathrm{sinh}(x)$, if $\Omk < 0$. And $h(z)$ is the normalized Hubble function,
\begin{equation}
    h(z) = \frac{H(z)}{H_0} = \sqrt{\Omm (1 + z)^3 + \OmL - \Omk (1 + z)^2}.
    \label{eq:h}
\end{equation}
We will also use quantities such as energy distance and luminosity distance
\begin{equation}
    d_E(z) = \sqrt{(1 + z)}\,l(z);\quad\quad d_L(z) = (1 + z) l(z).
    \label{eq:del}
\end{equation}
One of the main quantities, considered in this work, is the luminosity distance modulus
\begin{equation}
    \mu(z) = m_L - M_L = 5\log \bigg(\frac{d_L(z)}{1\,\textrm{Mpc}}\bigg) + 25.
    \label{eq:mu}
\end{equation}
The HD for some observed objects we will call the set of points $(z_i, \mu_i)$. Having constructed a HD based on some catalogue of observed objects, we can estimate the cosmological parameters $(H_0, \Omm, \OmL, \Omk)$ using approximation procedures.

\section{Observational data}

As observational data, we must choose such objects for which the redshift and the distance modulus are measured independently of each other. Examples of such objects are type Ia supernovae, which are standard candles in cosmology.
However, it is not supernovae that are of greatest interest to us, but LGRBs: in this paper, we consider the possibility of using the latter as standard candles. For further analysis, we need the initial LGRB observational data, from which we will search for their distance moduli.

\subsection{Catalogue of supernovae used}

We have used the Ia Pantheon supernova catalogue. It contains distance moduli and their errors, as well as redshifts for 1\,048 supernovae~\footnote{\url{https://archive.stsci.edu/prepds/ps1cosmo/index.html}}. The redshift is determined in two different ways: in the heliocentric system $z_{\mathrm{hel}}$ and taking into account the motion of Galaxy relative to the cosmic microwave background radiation $z_{\mathrm{cmb}}$. In our work, it was the latter values that were used.

\subsection{Catalogue of gamma-ray bursts used}
\label{subsec:GRBcat}

% выборка на 1 июня 2021
We have used data from 174 LGRBs observed by the Swift space observatory~\footnote{\url{https://swift.gsfc.nasa.gov/archive/grb_table/}}. 
For each LGRB we used the following four values.
\begin{itemize}
    \item $z$ is redshift. It is determined by identifying a LGRB with its host galaxy. The (spectroscopic) redshift error is absent in the catalogue. It is considered as negligibly small, so we do not use it in our calculations.    \item $\Sobs$ is the LGRB observable fluence in $10^{-7}\,\text{erg}/\text{cm}^2$. The catalogue for these values also presents the errors corresponding to the confidence interval with the 90\% error level (approximately 1.6$\sigma$). It is important for us that the errors must correspond to the $1\sigma$ level, so these errors were divided by 1.6.
    \item $\Ep$ is the LGRB peak energy in keV. This is the energy value at which the $E^2N(E)$ spectrum peak is observed. The peak energy errors are asymmetric, i.e. the upper and lower values are different. Moreover, these errors are only determined for approximately 1/3 of LGRBs, and they are missed for the remaining 2/3 of LGRBs.
    \item $\alpha$ is the dimensionless parameter to describe shape of the LGRB spectrum according to the CPL (cut-off power law) model, given by Equation~(\ref{eq:CPL}). The errors for this parameter are also asymmetric, and they are given for almost all gamma-ray bursts.
\end{itemize}

As mentioned above, the errors of some quantities are not presented for all available LGRBs in our sample. In order to take into account points without errors along with points for which errors are presented, we used the following solution: all unknown errors were replaced with median relative errors for a given quantity in our sample. The original and enhanced catalogues can be found in Appendices~\ref{app:B} and~\ref{app:C}, respectively.

\section{Gamma-ray bursts}
\subsection{Photometric quantities of LGRBs}
\label{subsec:GRBphot}

These formulas (with the exception of the CPL spectrum model) are described in more detail in~\citet{shirokov2020high}. The CPL spectrum model is defined as follows:
\begin{equation}
    N(E)\bigg[ \frac{\text{photons}}{\text{keV}\,\text{c}\,\text{sm}^2} \bigg] = AE^\alpha \exp\bigg(-\frac{E}{\Ep} (2+\alpha)\bigg),
    \label{eq:CPL}
\end{equation}
where $A$ is a normalisation parameter. This model is used for converting from the observed fluence to the standard ``bolometric'' gamma fluence (from 1 to 10$^4$ keV)
\begin{equation}
    \Sbolo = \Sobs \frac{\int_{\frac{1}{1+z}}^{\frac{10^4}{1+z}} E\, N(E) \, \di E}{\int_{15}^{150} E\, N(E) \, \di E},
    \label{eq:integralspectr}
\end{equation}
in the integration limits, the energies are given in keV. The integration limits in the denominator correspond to the spectral range of the main telescope (BAT) of the Swift space observatory (from 15 to 150 keV). 

We also want to introduce a phenomenological model for taking into account the Malmquist effect and gravitational lensing in accordance with~\citet{shirokov2020high}. It is carried out in the form of a parameterisation using a following power law:

\begin{equation}
    \Sbolocor = \frac{\Sbolo}{(1+z)^k}.
    \label{eq:correction}
\end{equation}
The parameter $k$ characterises the change in the bolometric fluence due to the Malmquist effect and gravitational lensing.

Knowing the bolometric fluence, we can find the isotropic equivalent radiated energy
\begin{equation}
    \Eiso = 4\pi d_E^2\, \Sbolocor.
    \label{eq:Eiso}
\end{equation}
This value has the following meaning: it is the energy that would be emitted by a LGRB if it emitted energy isotropically (with a spherical indicatrix).

Another quantity we need is the peak energy in the source frame. It is equal to
\begin{equation}
    \Epi = \Ep (1 + z).
    \label{eq:Ep}
\end{equation}

\subsection{The Amati relation and the extended Amati relation}

Lorenzo Amati was one of the first who discovered the correlation between the LGRB spectrum peak energy in the source frame $\Epi$, given by Equation~(\ref{eq:Ep}), and the LGRB isotropic equivalent radiated energy $\Eiso$, given by Equation~(\ref{eq:Eiso})~\citep{Amati2002}. This dependency is called the Amati relation and can be written in this form:
\begin{equation}
    \log \frac{\Eiso}{1\,\text{erg}} = a\,\log \frac{\Epi}{1\,\text{keV}} + b.
\label{eq:Amatirel}
\end{equation}
The coefficients $a$ and $b$ are called Amati parameters. Equation (\ref{eq:Amatirel}) can be rewritten as follows:
\begin{equation}
    \frac{\Eiso/\text{1 erg}}{(\Epi/\text{1 keV})^a} = 10^b = \mathrm{const}.
    \label{eq:Amatirelphyssense}
\end{equation}
Thus, for each LGRB, the ratio $\Eiso / {\Epi}^a$ is the same with the same accuracy as the Amati relation~(\ref{eq:Amatirel}). Based on the presence of such a constant quantity for LGRBs, it follows that they could be employed as standard candles, similar to type Ia supernovae.

It follows from this that if the Amati parameters $(a, b)$ are known, then the LGRB distance moduli $\mu$ can be calculated regardless of the cosmological model using the Amati relation. Indeed, substituting the expression for the isotropic equivalent radiated energy $\Eiso$ with the corrected bolometric fluence, given by Equation~(\ref{eq:correction}), into the Amati relation, one can express from there the energy distance $d_E$ (Equation~(\ref{eq:del})) and convert it into the distance modulus $\mu$ (Equation~(\ref{eq:mu})). This will result in the following expression:
\begin{equation}
    \muagrb = 25 + \frac{5}{2}\bigg[ \log \frac{(z+1)^{k+1}}{4\pi \Sbolo} + a\log (\Epi) + b \bigg].
    \label{eq:Amatiext}
\end{equation}
We will call this expression the extended Amati relation.

In accordance with equations (\ref{eq:metricdist}), (\ref{eq:del} for $D_L$), and (\ref{eq:mu}), the distance modulus is given by
\begin{equation}
\mu(z) = 25 + 5\log \left(\frac{c}{H_0}\right) + f(z, \Omm, \OmL, k),
\end{equation}
where $f$ represents the corresponding cosmological model derived function. So, the distance scale depend on the value of the Hubble constant $H_0$. The Pantheon SNIa catalogue uses $H_0=70$ km/s/Mpc to determine the absolute magnitudes of supernovae $M$. In the subsequent part of the article, we verify our methods by varying the parameters of the cosmological model, including $H_0$
(the correctness of the method implies that the obtained values of $H_0$ should be equal to the initial one within the margin of error). 
%However, it is important to note that we do not claim any attempt to define the value of $H_0$ or the distance scale itself. 
Therefore, the distance scale and all the values of the Amati parameters obtained in this paper are tied $H_0=70$ km/s/Mpc. Changing this value will, in particular, result in a change in the values of $a$ and $b$.
% Despite the fact that in the subsequent part of the article we verify our methods by varying cosmological model parameters including $H_0$, it is not an attempt to address the Hubble tension problem. 

% почему Н_0 где-то не 70?
% здесь надо написать, что фиксируем Н_0
% и где и на каком основании/с какой целью и в каком смысле варьируем H_0
% сослаться на формулы выше (16) и (17) и сказать что b связано с 5 log (c/H_0)

\section{Description of tasks and stages of their solution}\label{sec:tasks}
\label{sec:decompose}

\subsection{Direct problem}\label{sec:direct}

We will divide the problem-solving process into separate stages, providing a detailed description for each of them.

In order to solve the direct problem, one need to determine the parameters of the $\Lambda$CDM model using the provided LGRB sample by following stages.

\begin{enumerate}
    \item \textit{Obtaining initial data.} At the initial stage, we have the Swift LGRB catalogue, described in detail in Section~\ref{subsec:GRBcat}. It is a table of the following form as in Table~\ref{tab:GRBcatalogue}.
    Each LGRB in this table is described by a set of parameters $(z, \alpha, \Ep, \Sobs)$. The parameters $\alpha$ and $\Ep$ have asymmetric errors. The symbols $\Deltal$ and $\Deltau$ mean the ``bottom'' and ``top'' errors, respectively. The symbol $\Delta_{90\%}$ means the error corresponding to the 90\% confidence interval for a normal distribution. To bring this error to a standard deviation (confidence interval 68.3\%), one need to divide it by a factor $1.6 \approx \Phi^{-1}\big((1+0.9 )/2\big)$, where $\Phi$ is the integral function of the standard normal distribution and $\Phi^{-1}$ is its inverse function (quantile). As mentioned in Section~\ref{subsec:GRBcat}, the LGRB catalogue is incomplete, because some errors are missing. In total, there are 174 LGRBs in our sample.
    \item \textit{Extending the catalogue.} We decided to fill the missing errors in the original table by calculating the median for the relative errors of the corresponding parameter. After these manipulations, we got the following sample in Table~\ref{tab:GRBsample}.
    \item \textit{Calculation of the quantities $\Eiso$ and $\Epi$.} Using some given value for the parameter $k$ (For example, $k=0,\,0.5,\,1$), we can use the formulas from Section~\ref{subsec:GRBphot} to calculate the isotropic equivalent radiated energy $\Eiso$, given by Equation~(\ref{eq:Eiso}), and the peak energy in the source frame $\Epi$, given by Equation~(\ref{eq:Ep}), for each LGRB.
    \item \textit{Computing the Amati parameters.} Using some of the linear regression methods, we can obtain an estimate for the parameters $(a, b)$, which are the usual parameters of a linear regression in the Amati relation~(\ref{eq:Amatirel}).
    \item \textit{Calculation of gamma-ray burst distance modules.} Given the parameter $k$ and the computed Amati parameters $(a, b)$, we can find the LGRB distance modulus $\muagrb$ using the extended Amati relation~(\ref{eq:Amatiext}).
    \item \textit{Constructing a HD.} With the $\muagrb$ values we calculated above and with the $z$-source catalogue, we can construct a redshift--distance modulus diagram, also called a HD.
    \item \textit{Estimating the parameters of the $\Lambda$CDM model.} Using approximation algorithms, one can approximate the theoretical dependence $\mu(z)$ (Equation~(\ref{eq:mu})) depending on the parameters of the cosmological $\Lambda$CDM model by using a some initial sample, for instance, the SN Ia supernovae catalogue to the HD constructed for our LGRBs.
\end{enumerate}

    \begin{table*}
        \centering
        \begin{tabular}{l||c|ccc|ccc|cc}
        \hline
            LGRB number & $z$ & $\alpha$ & $\Deltal \alpha$ & $\Deltau \alpha$ & $\Ep$ & $\Deltal \Ep$ & $\Deltau \Ep$ & $\Sobs$ & $\Delta_{90\%} \Sobs$ \\ \hline
            GRB... & ... & ... & n/a & ... & ... & ... & n/a & ... & ... \\
            GRB... & ... & ... & ... & n/a & ... & n/a & ... & ... & ... \\
        \hline
        \end{tabular}
        \caption{Form of the original catalogue of LGRBs. The full catalogue is presented in Appendix~\ref{app:B}.}
        \label{tab:GRBcatalogue}
    \end{table*}
    
    \begin{table*}
        \centering
        \begin{tabular}{l||c|ccc|ccc|cc}
        \hline
            LGRB number & $z$ & $\alpha$ & $\Deltal \alpha$ & $\Deltau \alpha$ & $\Ep$ & $\Deltal \Ep$ & $\Deltau \Ep$ & $\Sobs$ & $\Delta \Sobs$ \\ \hline
            GRB... & ... & ... & ... & ... & ... & ... & ... & ... & ... \\
            GRB... & ... & ... & ... & ... & ... & ... & ... & ... & ... \\
        \hline
        \end{tabular}
        \caption{Form of the extended catalogue of LGRBs. $\Sobs$ errors are reduced to $1\sigma$. The full catalogue is presented in Appendix~\ref{app:C}.}
        \label{tab:GRBsample}
    \end{table*}

But this problem has an issue when passing from point 2 to point 3. The fact is that Equation~(\ref{eq:Eiso}) for $\Eiso$ contains the distance $d_E$, which, in turn, depends on the choice cosmological model, what can be seen from Equations (\ref{eq:metricdist}, \ref{eq:h}, \ref{eq:del}). Because of this, it turns out that the Amati parameters $(a, b)$, and, as a consequence, the distance moduli $\muagrb$ of LGRBs are also dependent on the cosmological model!

As a result, the parameters of the $\Lambda$CDM model $(H_0, \Omm, \OmL, \Omk)$ that we get at the output also depend on the chosen dependency $\mu(z)$ and, as a consequence, on some initially chosen cosmological parameters $(H_0, \Omm, \OmL, \Omk)_0$. This is known as a circularity problem~\citep{Kodama2008}.

\subsection{Addressing the circularity problem}
    
There are several approaches to resolve the circularity problem. The first approach based on using known standard candles such as supernovae, globular clusters or others to calibrate near GRBs as a standard candles. In particular, in~\citet{Lovyagin2022} we used approximation of the dependence $\mu(z)$ by the HD via type Ia supernovae from the Pantheon catalogue. 
The second approach is to simultaneously constrain the calibration parameters and the cosmological parameters by considering a chosen likelihood function, analogically \citet{amati2013, demianski2021prospects}.

We also suggest here another approach based on method of iterations. Thus, the parameters $(H_0, \Omm, \OmL, \Omk)_i$ obtained at the output of the algorithm can be used in the repeated solution when passing between point 2 and point 3, and as a result, new estimates for these parameters can be obtained $(H_0, \Omm, \OmL, \Omk)_{i+1}$, and so on until convergence is reached. 

\begin{figure}
    \centering
    \includegraphics[width=0.4\textwidth]{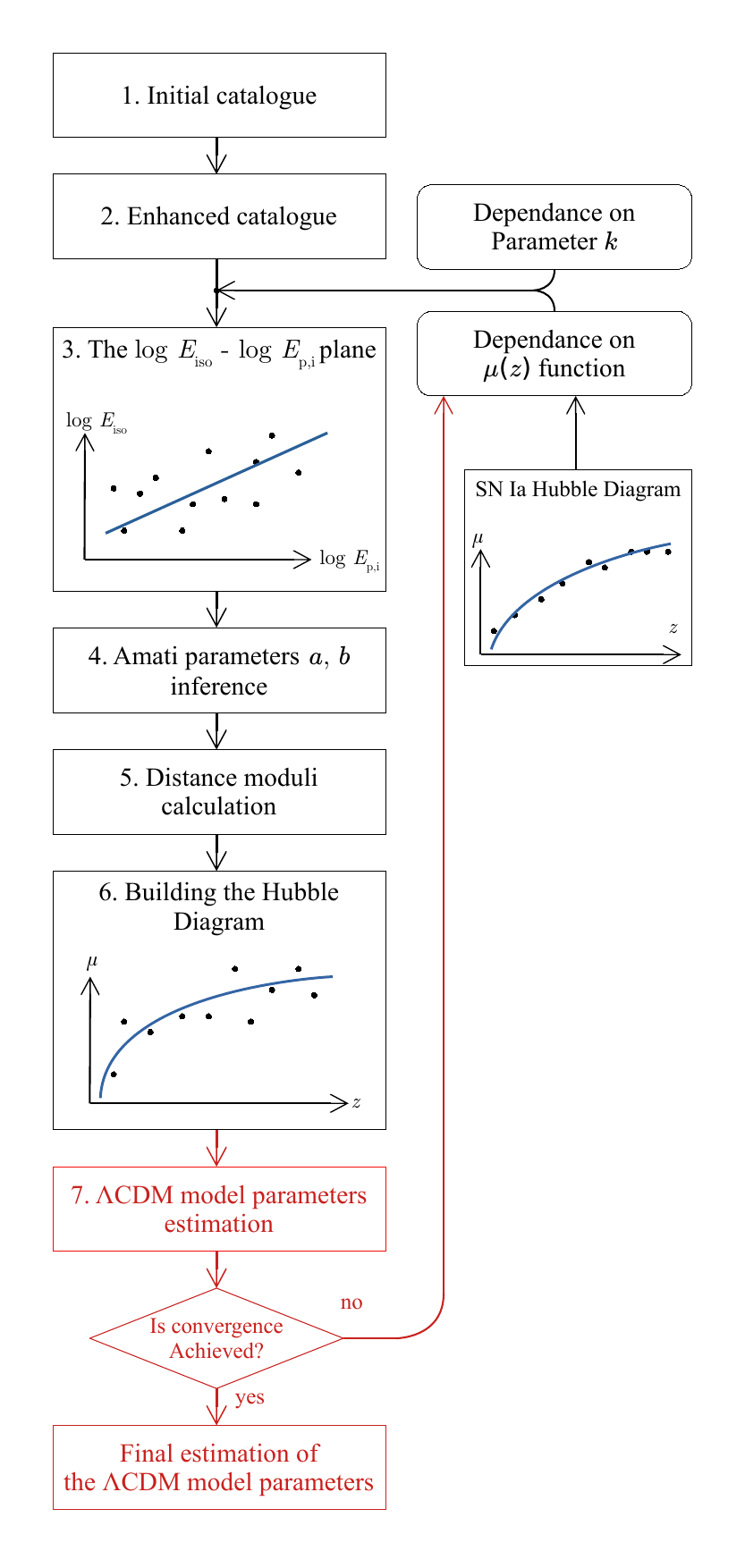}
    \caption{Visualisation of the direct problem and its circularity. The scheme represents the process of deriving HD from the SN-derived $\Lambda$CDM model parameters (black). It serves as an illustration within this study, however, this approach suffers from a dead-loop (circularity problem) and cannot be used as a cosmological model parameter test. A potential solution using iterative methods is highlighted in red. In this study, it is stated as a task for a possible future research; instead, a solution to the inverse problem, illustrated in Figure \ref{fig:scheme2}, has been employed.}

    \label{fig:scheme1}
\end{figure}

While we assume that this approach remains robust and accurate within the framework of Bayesian statistics, we refrain from employing it in this article, leaving it as a potential future research. Instead, the article addresses the inverse cosmological calibration problem as described in Section \ref{sec:iccp}, i.e., obtaining the Amati parameter values within the framework of the $\Lambda$CDM model with the gravitational lensing and Malmquist bias correction~\citep{shirokov2020high}, but not the parameters of the cosmological model itself. When addressing the direct problem, our focus is limited to constructing the HD (item \textit{vi} in \ref{sec:direct}) for illustrative purposes. The scheme of the direct problem solution approach for is illustrated in Figure \ref{fig:scheme1}: the completed part of the work is highlighted in black, while the possible addressing the circularity problem within this context is shown in red.

\subsection{Inverse cosmological calibration problem}\label{sec:iccp}
\label{sec:ICCP}

The core of the idea of the inverse cosmological calibration problem (ICCP) is that we use the least-squares algorithms to minimise the residuals from the HD ($\mu_{\mathrm{model}} - \mu_{\mathrm{GRB}}^{\mathrm{A}}$), while the points $\mu_i$ are calculated from the $\log \Epi$ -- $\log \Eiso$ plot (the Amati plane) that is varied by ($a,\,b,\,k$). In this study, we fix parameters $(H_0, \Omm, \OmL, \Omk)$ of the $\Lambda$CDM model, so optimal values for the parameters $a$, $b$, and $k$ are determined using this method.

Our main objective is to solve the ICCP, which involves determining the parameters of the Amati parameters $(a, b)$ and correction parameter $k$ using the provided LGRB sample by following stages.

\begin{enumerate}
    \item Using \textit{obtained initial data} and \textit{extended catalogue} as described in \ref{sec:direct}.
    \item \textit{Obtaining $\Lambda$CDM parameters}. This can be done, for example, using the Ia supernovae of the Pantheon catalogue as an input calibration sample (since the ICCP can be considered as another way to calibrate LGRBs from supernovae, it can address the circularity problem).
    \item \textit{Determining parameters} using least squares to minimise residuals of the LGRB HD. In current research we fix cosmological model parameter determined at the second stage. In the first case, we also fix the $k$-parameter to be equal to 0 and determine the values of $a$ and $b$. In the second case we determine all three values $a$, $b$ and $k$.
\end{enumerate}    

\begin{figure}
    \centering
    \includegraphics[width=0.49\textwidth]{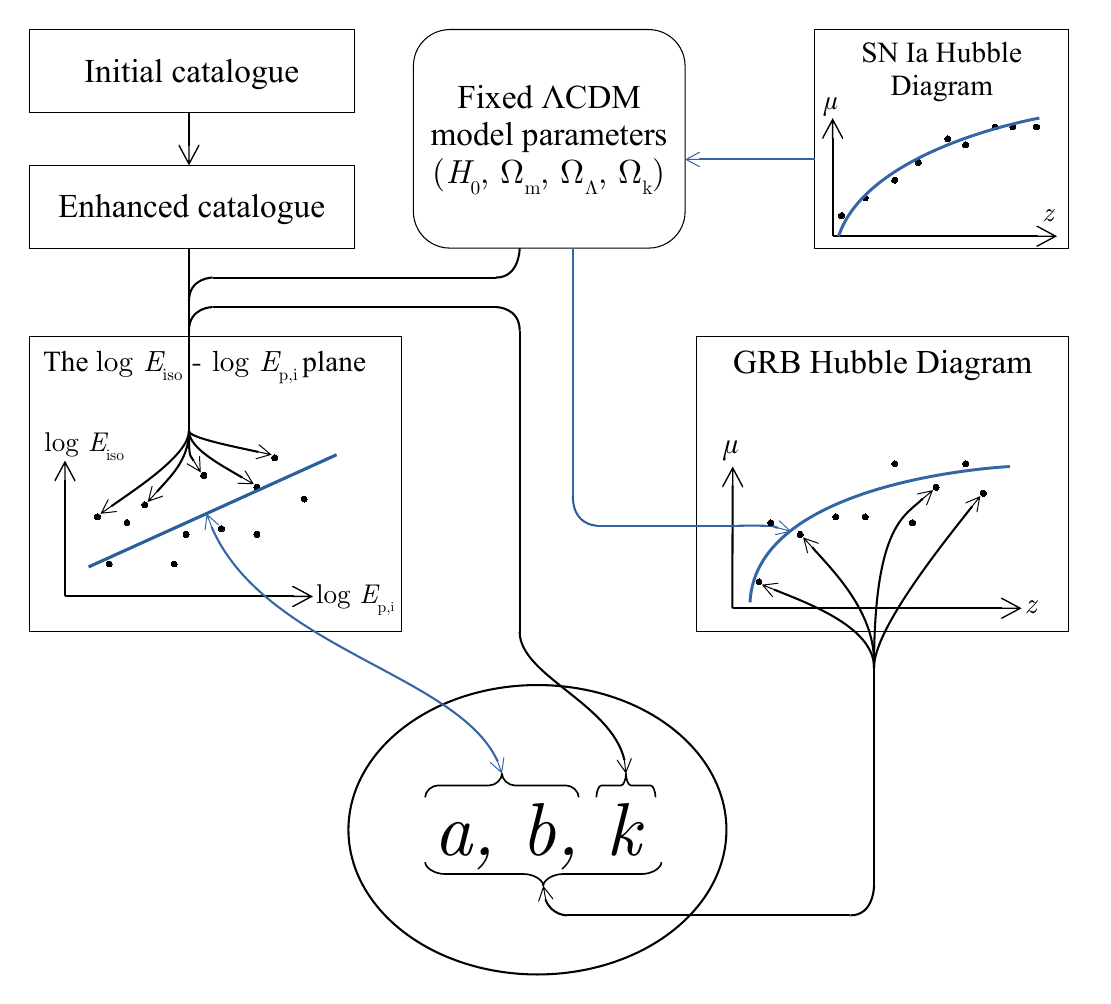}
    \caption{Scheme for solving the inverse problem. The influence of data and parameters on the calculated values is illustrated here. From the Pantheon type Ia supernova HD (top right), we determine the $(H_0, \Omm, \OmL, \Omk)$ parameters of the $\Lambda$CDM model, which we then assume to be fixed. Parameters $(a, b, k)$ are required. To estimate them, we minimise the deviations in the HD for LGRBs (right). The fixed parameters of the $\Lambda$CDM model define the dependence of the distance on the redshift $\mu(z)$, which is shown by the blue curve in the HDs. To find $\Eiso$, one need to use the formulas from Sections \ref{sec:cosm} and \ref{subsec:GRBphot}. These formulas depend on the choice of $k$ (Equation~(\ref{eq:correction})) and on the dependency of $\mu(z)$ (Equations~(\ref{eq:Eiso}, \ref{eq:del})). Therefore, the value of $\Eiso$ also depends on $k$ and fixed parameters of the $\Lambda$CDM model $(H_0, \Omm, \OmL, \Omk)$. The Amati parameters $a, b$ define the position of the straight line in the plane $\log \Eiso$ -- $\log \Epi$, which we called the $\log \Eiso$ -- $\log \Epi$ plane. In this case, all three sought-for parameters $a, b, k$ affect the LGRB distance moduli calculated using the extended Amati relation~\ref{eq:Amatiext}.}
    \label{fig:scheme2}
\end{figure}

A detailed scheme for solving the inverse problem is shown in Figure~\ref{fig:scheme2}.

In this approach, the parameter $k$ is considered as a correction parameter to the $\Lambda$CDM model. It is used as a correction for gravitational lensing and the Malmquist bias (GLMB-correction from \citep{shirokov2020high}). Determining it by solving the direct problem without addressing the circularity problem is not possible, whereas the inverse problem is free from this flaw. Assuming the accurate determination of the fundamental parameters of the $\Lambda$CDM model, fixing $k=0$ should yield the Amati parameter values $a$ and $b$ within the uncertainty limits that match those obtained from solving the direct problem without addressing the circularity problem.

\section{Methods}
\label{sec:methods}
\subsection{Approximation and regression algorithms used}
\label{subsec:regr}

As evident from Section \ref{sec:tasks} various stages required the usage of both linear regression (such as in calculating the Amati parameters) and non-linear regression (for example, in estimating the parameters of the $\Lambda$CDM model using the HD).

As an algorithm for linear regression, we have chosen the Theil-Sen method \citep{gilbert1987statistical}. The essence of this method is that for a given set of points $\{(x_i,y_i)\}_i$ the estimated slope $k$ of the linear function $y = kx + m$ is determined as the median value of all possible slopes $(y_i - y_j )/(x_i - x_j)$ for each pair of points. The coefficient $m$ of the linear relationship can then be estimated as the median of the values $y_i - kx_i$. The main advantage of this method over the classical method of least squares is its robustness, i.e. resistance to outliers.
% outliers -- выбросы

In the context of non-linear regression, we employed the \texttt{least\_squares} procedure, which is implemented in the \texttt{scipy.optimize} package of the Python programming language. This routine employs a technique known as the ``trust region reflective'' method to identify the minimum of the variance function as follows:
\begin{equation}
   F(\boldsymbol{\theta}) = \frac{1}{2}\sum\limits_{i} \rho(f_i^2(\boldsymbol{\theta})),
\end{equation}
where $\boldsymbol{\theta}$ represents the vector of estimated parameters, $f_i$ denotes the model's deviation from the data, and $\rho$ denotes the loss function. Choice for the identity mapping as the loss function $\rho$ simplifies the problem into the classical finding of the minimum sum of squared deviations. However, this approach can introduce significant bias in parameter estimates due to data outliers. To mitigate the impact of outliers, an alternative loss function must be selected. For instance:
\begin{equation}
\rho(f^2) = 2C^2 \left(\sqrt{1 + f^2 / C^2} - 1\right),
\end{equation}
where $C$ serves as a conditional soft boundary, separating outliers. For the value of this constant, we have chosen doubled the standard deviation of $f_i$.

\subsection{Monte Carlo method for calculating errors of indirect measurements used}
\label{subsec:montecarlo}

We chose the Monte Carlo method as the error propagation method \citep{anderson1976error,albert2020monte,gorokhov2023nonparametric}. To proceed to the description of the essence of this method, first one need to decide on the interpretation of the error. 

The standard approach for addressing the error propagation issue is referred to as the linear uncertainty propagation (LUP) theory. In this approach, errors are treated as standard deviations, and the values are regarded as the means. Thus, variable values along with their errors are determined using the normal distribution, which represents the potential range of the variable through distribution parameters. When a measured value $\mu$ is provided with an error of $\sigma$, it signifies that this value is random and follows a normal distribution with a mean of $\mu$ and a standard deviation of $\sigma$.

The standard formula for calculating the error of indirect measurement is derived from simple considerations. Let us have $n$ normally distributed random variables with known expected values $\{x_i\}_{i=1}^n$, variances $\{\sigma_i^2\}_{i=1}^n$ and covariances $\{\sigma_{ij}\}_{i \neq j}$. Let $f$ also be a linear function of these quantities:
\begin{equation}
    f(x_1,...x_n) = \sum\limits_{i=1}^n a_i x_i.
\end{equation}
The value of $f$ will also have a normal distribution. Its variance can be obtained as follows:
\begin{equation}
\begin{aligned}
    \sigma_f^2 & = \mathrm{E}\big[ (f - \mathrm{E}f)^2 \big] = \mathrm{E}\bigg[ \Big(\sum\limits_{i=1}^n a_i(x_i - \mathrm{E} x_i)\Big)^2 \bigg] = \\
    & = \sum\limits_{i=1}^n a_i^2 \mathrm{E} \Big[ (x_i - \mathrm{E}x_i)^2 \Big] + 2\sum\limits_{i\neq j} a_i a_j \mathrm{E} \big[ (x_i - \mathrm{E} x_i)(x_j - \mathrm{E} x_j) \big] = \\
    & = \sum\limits_{i=1}^n a_i^2 \sigma_i^2 + 2\sum\limits_{i\neq j} a_i a_j \sigma_{ij},
\end{aligned}
\end{equation}
or, if we treat the standard deviation as an error,
\begin{equation}
    \Delta f = \sqrt{a_1^2\Delta x_1^2 + ... + a_n^2\Delta x_n^2 + 2 a_1 a_2 \mathrm{cov}(x_1,x_2) + ...} \, .
\end{equation}

If the function $f$ is non-linear, it is represented as an expansion in a Taylor series up to the first order:
\begin{equation}
\label{eq:unpropclassic}
    \Delta f = \sqrt{\bigg( \frac{\partial f}{\partial x_1} \bigg)^2 \Delta x_1^2 + ... + \bigg( \frac{\partial f}{\partial x_n} \bigg)^2 \Delta x_n^2 + 2 \frac{\partial f}{\partial x_1} \frac{\partial f}{\partial x_2} \mathrm{cov}(x_1,x_2) + ...} \, .
\end{equation}
This is the standard formula for calculating the error of indirect measurements. Unfortunately, this approach has limited applicability due to the following disadvantages:
\begin{itemize}
    \item Measured values and errors are treated as expected values and standard deviations of random variables that follow a normal distribution. Therefore, this method is not applicable to data that has asymmetric errors.
    \item It is necessary that the errors be small enough and the functions smooth enough that the inaccuracy due to representing the function in a linearised form is negligible.
\end{itemize}
In our case, the observational data exhibit asymmetric errors with large values. Also, the data transformation function $f$ is far from linear.
In addition, it cannot be guaranteed that the indirect (i.e., transformed) values will obey the normal distribution. This can be demonstrated with a simple example.

Let two quantities with errors be given: $x\pm \sigma_x$ and $y\pm \sigma_{y}$. We want to find their ratio $x/y$ and its error $\sigma_{x/y}$. As is known, the value, which is the ratio of two normally distributed random variables, obeys the Cauchy distribution (among astrophysicists it is better known as the Lorentz profile). The latter is a standard example of a distribution for which neither the expected value, nor the variance, nor the higher-order moments are defined.
This example shows that for our case, not only the standard error propagation formula (\ref{eq:unpropclassic}) is inapplicable, but also the interpretation of values with an error as random variables following the normal distribution with specified mean and variance is also not valid. Therefore, it is necessary to choose another interpretation of the errors and error propagation method, which would be a generalisation to arbitrary distributions.

The concepts of mean and standard deviation can be naturally generalised using quantiles. The $0.5$ level quantile (median) is defined for any continuous distribution, and in the case of a normal distribution, it coincides with the expected value (mean). The quantiles $0.16$ and $0.84$ are the bounds of the confidence interval with the confidence level $0.68$, which in the case of a normal distribution reduces to the interval with bounds $(\mu-\sigma,\mu+\sigma)$. Thus, for arbitrary distributions, we will use the median as a ``generalised mean'' instead of the mean, and the differences of the $0.84$ level quantile with the median ($q_{0.84} - q_{ 0.5}$) and the median with the $0.16$ level quantile ($q_{0.5} - q_{0.16}$) as the top and bottom errors, respectively.

We can now move on to describing how to calculate indirect measurement (i.e., transformed values) errors using the Monte Carlo method. 
The concept is as follows: given that each input value is represented by a measured value and its error value, which define a form of continuous distribution of a random variable (typically a normal distribution in the simplest scenario), we can simulate this distribution using a random number generator. This simulation involves generating a sample of size $N$ that follows the appropriate distribution. That is, each of the input values is represented as a sample of size $N$. The problem can be represented as $N$ different realisations of our experiment, in each of which all indirect measurements are calculated. Thus, for each indirect value, we will also have a sample of size $N$, which will be a model of the distribution of this value as a random variable. Based on this sample, it is possible to determine the quantiles of the $0.16$, $0.5$ and $0.84$ levels in order to estimate the value of this value and its errors.

\begin{figure*}
    \centering
    \includegraphics[width=0.49\textwidth]{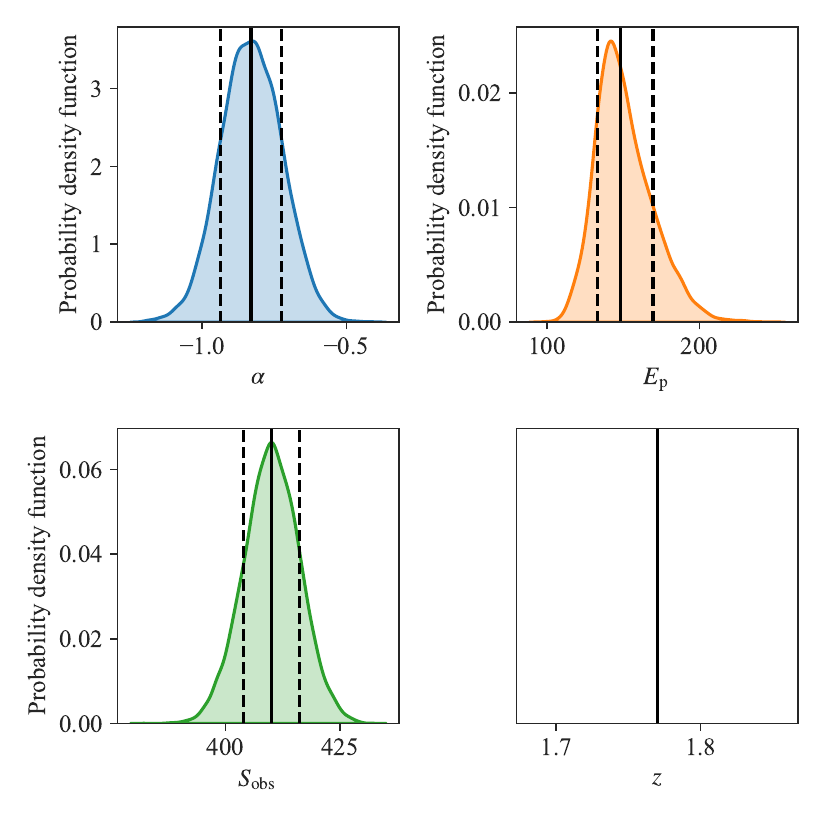}
    \vspace{0.3cm}
    \includegraphics[width=0.49\textwidth]{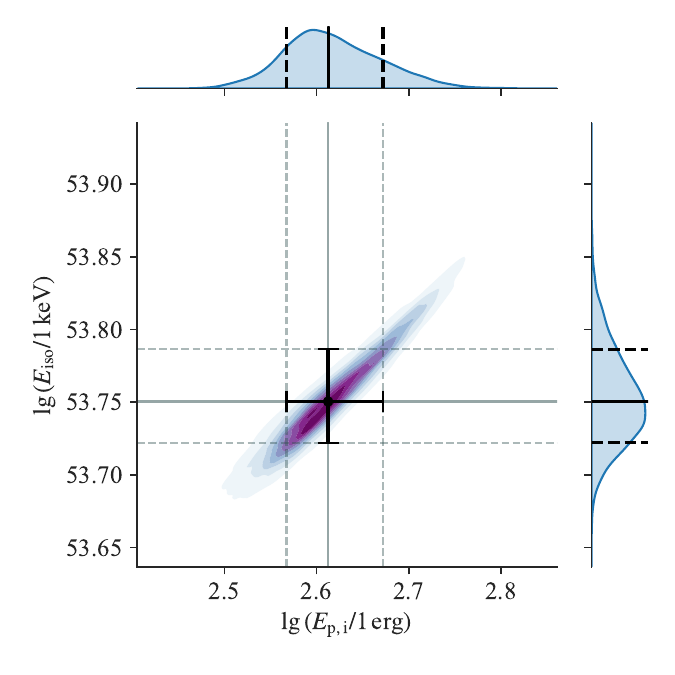}
    \caption{An example of the application of the Monte Carlo method for one of the LGRBs (GRB110422A). Upper block: original 4 parameters $\alpha$, $\Ep$, $\Sobs$, and $z$. For the first three, Monte Carlo distributions of size $N=10\,000$ are presented. The vertical solid black line marks the median, and the dotted line marks the errors. The redshift $z$ has no error and is therefore shown in the figure as a single vertical line. Bottom block: distributions of the logarithms of the parameters $\Epi$ and $\Eiso$ calculated using the Monte Carlo method. The solid lines also mark the positions of the medians, and the dotted lines mark the errors.}
    \label{fig:MC_example}
\end{figure*}

If the initial value has symmetrical errors, then the normal distribution can be used as a model one for it. For values that have different upper and lower errors, we used a distribution consisting of two halves of different Gaussian distributions. This is also the so-called split-normal distribution. An example of such a distribution, as well as an example of using the Monte Carlo method to estimate the errors of indirect measurements, is shown in Figure~\ref{fig:MC_example}. Another important point is that the value $\Ep$ (the peak energy of the LGRB spectrum) cannot be negative. 
Some of the $\Ep$ values from our LGRB data set had remarkably huge lower uncertainties, so that trying to create a Monte-Carlo sample could lead to negative $\Ep$ values, which has absolutely no physical sense. Because of that, in the case of $\Ep$ variables, we draw the split-normal distributions in the space of $\log \Ep$.
To prevent ``tails'' of distributions from falling into the negative area, we created Monte Carlo model samples not for the $\Ep$ value itself, but for its logarithm. So, the peak energy values are drawn using the log-split-normal distributions. Taking the logarithms does not move the quantiles, so the errors in this case are not changed. 

This method has a number of advantages. Firstly, it is easy to implement and interpret the results. Secondly, this method is universal in the sense that it is applicable to any kind of function. Thirdly, in this method, any correlations between random variables are automatically taken into account in subsequent calculations, without the need to calculate derivatives. The only significant disadvantage is the increase in calculation time by $N$ times (in our case $N=10\,000$), but we live in an era where we can afford it.

\section{Software implementation}

To solve the tasks, we wrote Python code. The corresponding repository is publicly available at \url{https://github.com/Roustique/sngrb}. The repository includes our software implementation of the following steps:

\begin{itemize}
    \item Estimating the parameters $(H_0, \Omm, \OmL, \Omk)$ of the $\Lambda$CDM model from the HD ${(z_i,\mu_i)}_i$ for a sample of observational objects.
    \item Solving the direct problem (see Section~\ref{sec:decompose}, Figure~\ref{fig:scheme1}) of constructing a HD for LGRBs.
    \item Solving the inverse problem (see Section~\ref{sec:decompose}, Figure~\ref{fig:scheme2}) of finding the Amati parameters $a, b$, and the correction parameter $k$ based on a fixed cosmological model.
\end{itemize}

The software implementation was done utilising the following libraries.
\begin{itemize}
    \item \texttt{NumPy} -- for supporting arrays, including multidimensional ones, and functions that operate on them.
    \item \texttt{SciPy} -- we used the \verb"optimize.least_squares" procedure for non-linear regression and \texttt{stats.mstats.theilslopes} for linear regression using the Theil-Sen method (details in Section~\ref{subsec:regr}). We also used the \texttt{stats} module for implementing cumulative and differential distribution functions, as well as quantiles.
    \item \texttt{Pandas} -- for working with catalogues.
    \item \texttt{Matplotlib}, \texttt{Seaborn}, \verb"mpl_scatter_density" -- for generating plots, graphs, and charts.
    \item \texttt{Joblib} -- for parallelization.
    \item \texttt{Numba} -- for just-in-time (jit) compilation.
\end{itemize}

\section{Results}
\subsection{Obtaining the $\Lambda$CDM model parameters from supernovae}\label{sec:snlcdm}

\begin{figure}
    \centering
    \includegraphics[width=0.49\textwidth]{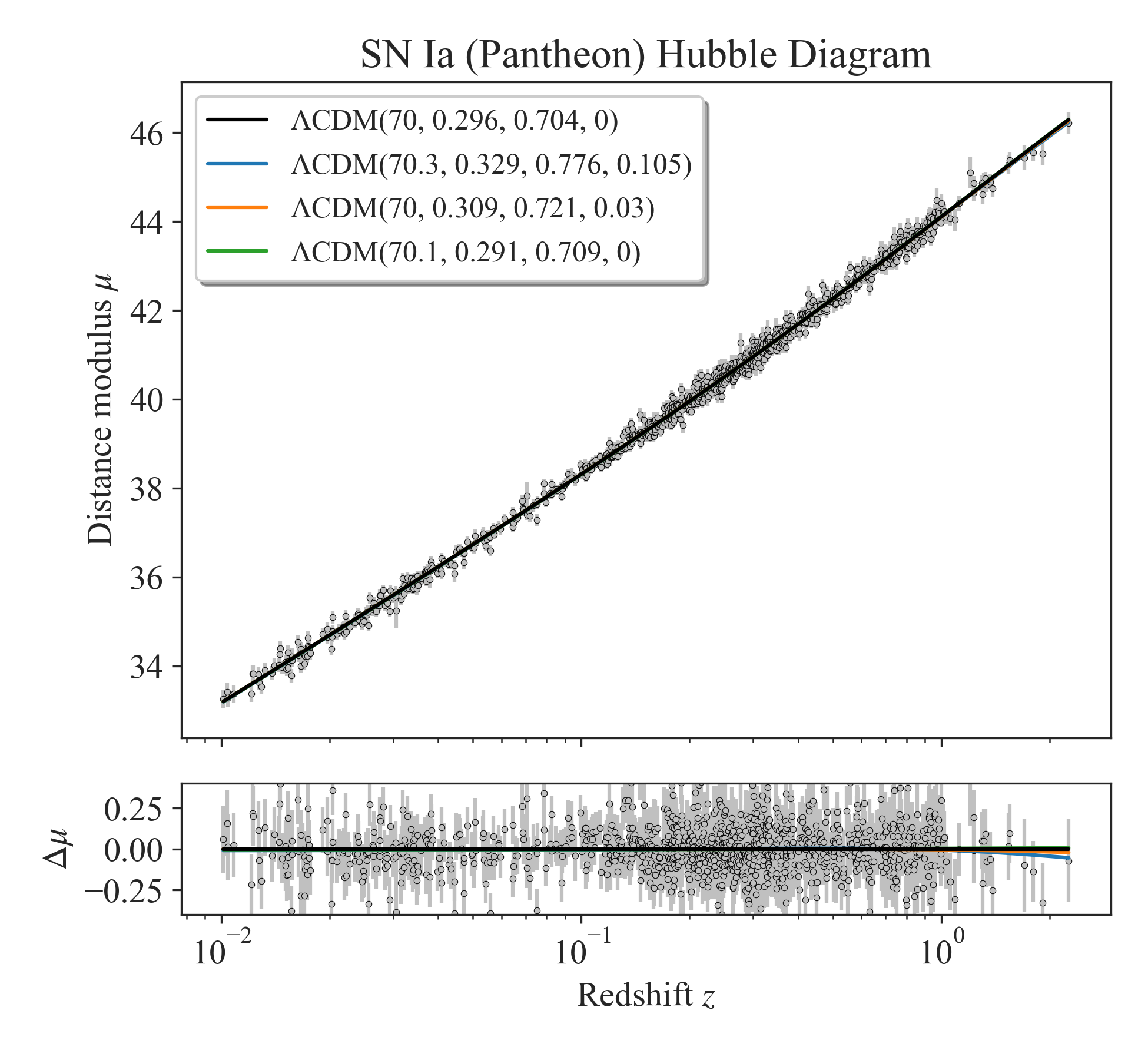}
    \caption{Curves corresponding to $\Lambda$CDM models with parameters found from supernova data. The blue line is the model with all free parameters. The orange line is the model with the fixed Hubble constant $H_0=70$. The green line is the model with a fixed curvature contribution $\Omk=0$. The black line is the model with fixed parameters $H_0=70, \Omk=0$.}
    \label{fig:HDSN}
\end{figure}

Using the procedure \texttt{least\_squares} with a loss function that is the identical mapping (written in detail in Section~\ref{subsec:regr}), we have estimated the parameters of the $\Lambda$CDM model from supernovae from the Pantheon catalogue. We have used 4 different models: in one, all 4 cosmological parameters $(H_0, \Omm, \OmL, \Omk)$ varied, and in the others, the parameters $H_0=70$ and/or $\Omk=0$ were fixed. The standard model is shown in Figure~\ref{fig:HDSN} with a black line and its parameters, along with the parameters of other models, are presented in Table~\ref{tab:HDSN}. The four cases considered allow us to conclude that our approximation algorithms work correctly, since the values of the cosmological parameters remain close to ($H_0=70$, $\Omm=0.3$, $\OmL=0.7$).

Since the Pantheon SN Ia catalogue data is known to be bound to $H_0=70$ and the $\Lambda$CDM model is bound to $\Omk=0$ we decided to choose a model with fixed values of this parameters as the standard. Other models were used to check the robustness of our methods. In this case we have obtained $\Omm=0.296$ and $\OmL=0.704$ as best-fitting values\footnote {These values differ from those determined, for example, in \citep{brout2022pantheon} and \citep{aghanim2020planck}, primarily due to fixing  $H_0=70$ and not relying on Cepheid or CMB measurements to derive the Hubble constant value.}.

\begin{table}
    \centering
    \begin{tabular}{r|cccc}
    \hline
         & $H_0$ & $\Omm$ & $\OmL$ & $\Omk$ \\ \hline
        fixed $H_0=70$ & 70 & 0.309 & 0.721 & 0.03 \\
        free parameters & 70.3 & 0.329 & 0.776 & 0.105 \\
        fixed $H_0=70$ and $\Omk=0$ & 70 & 0.296 & 0.704 & 0 \\
        fixed $\Omk=0$ & 70.1 & 0.291 & 0.709 & 0 \\ \hline
    \end{tabular}
    \caption{The estimated parameters of the $\Lambda$CDM model based on supernovae from the Pantheon catalogue. The model with fixed $H_0=70$ and $\Omk=0$ is chosen as the standard.}
    \label{tab:HDSN}
\end{table}

\subsection{Direct problem}
\label{subsec:rests}

\begin{figure}
    \centering
    \includegraphics[width=0.40\textwidth]{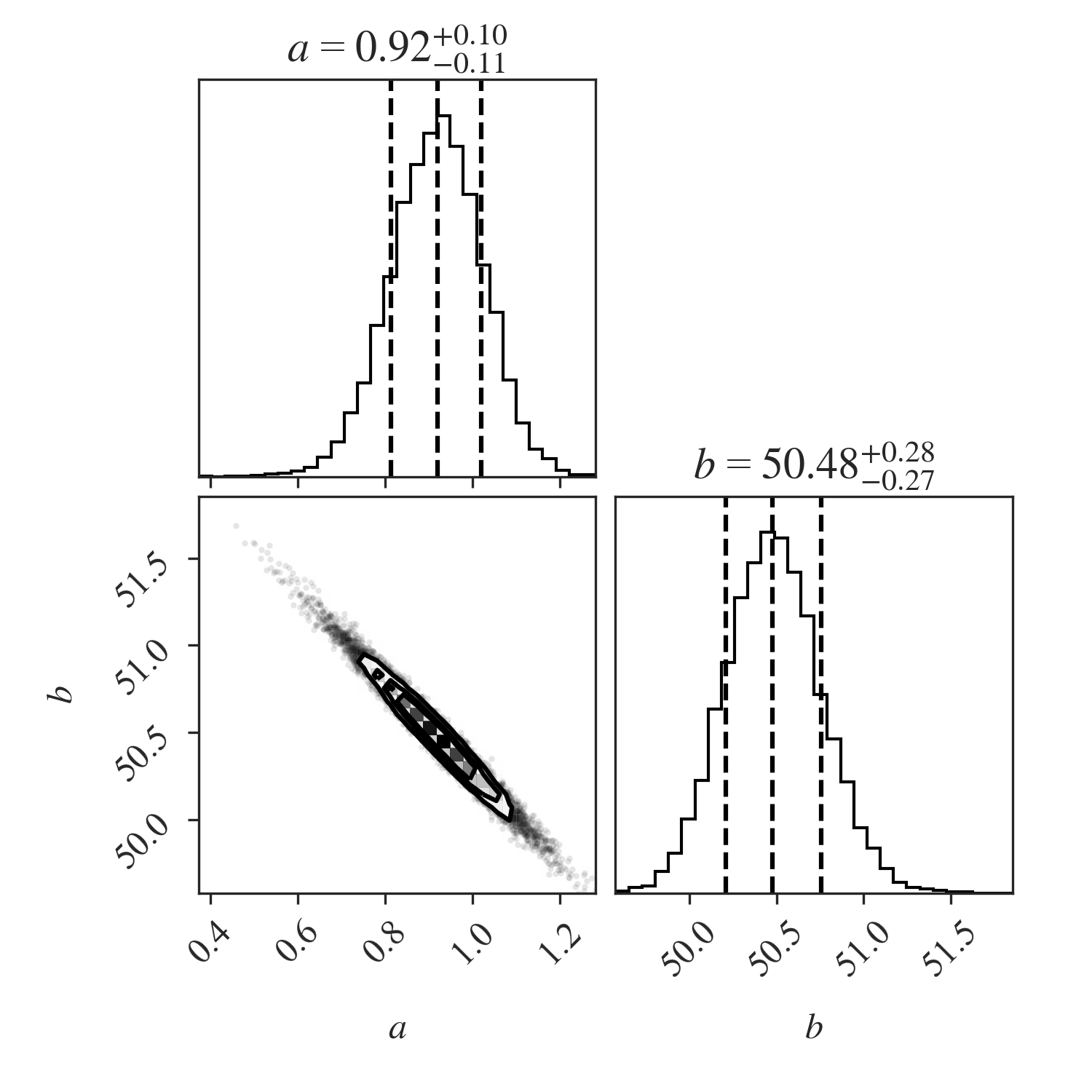}
    % \label{fig:cornerts}
    \includegraphics[width=0.48\textwidth]{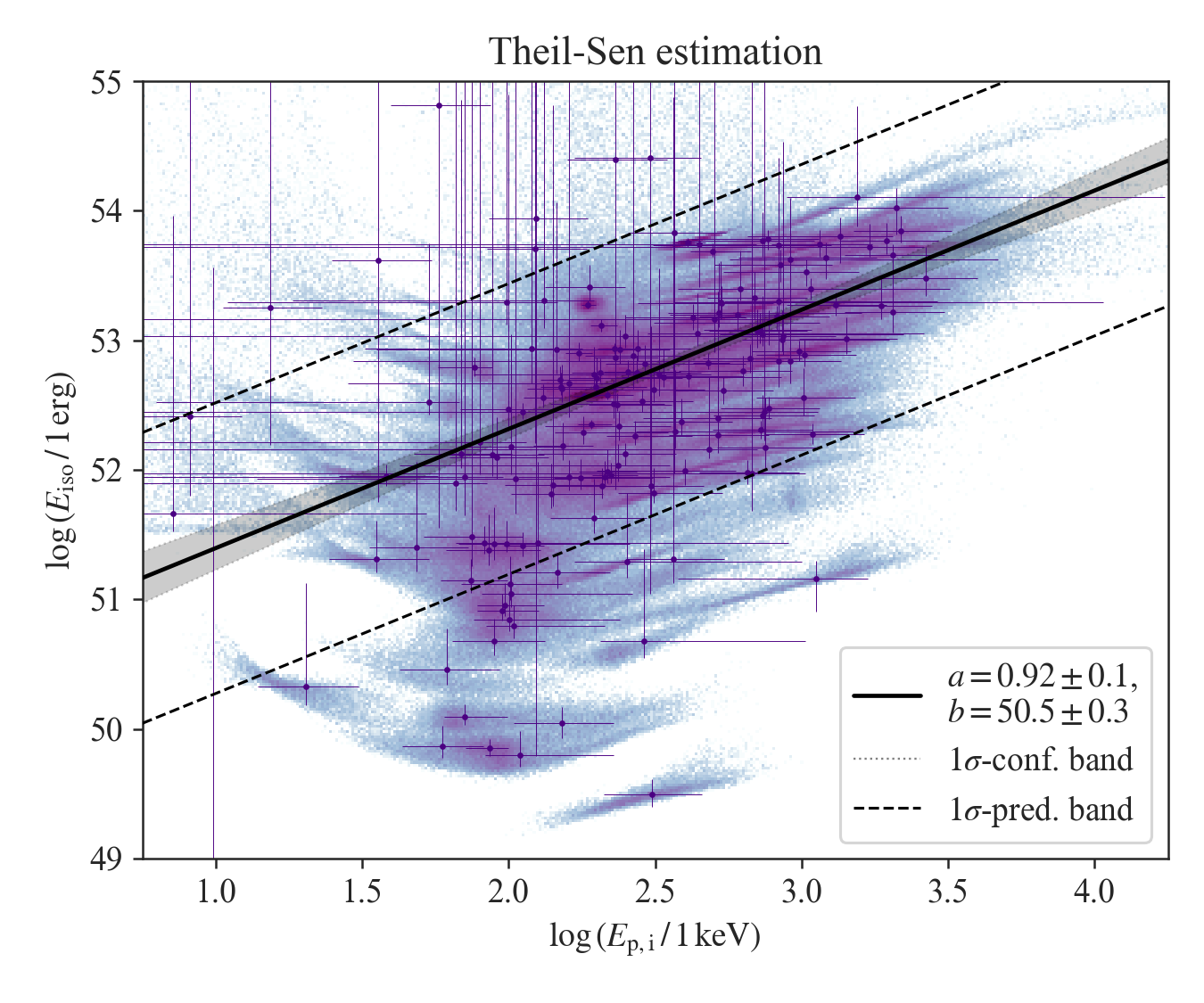}
    % \label{fig:Amatits}
    \includegraphics[width=0.48\textwidth]{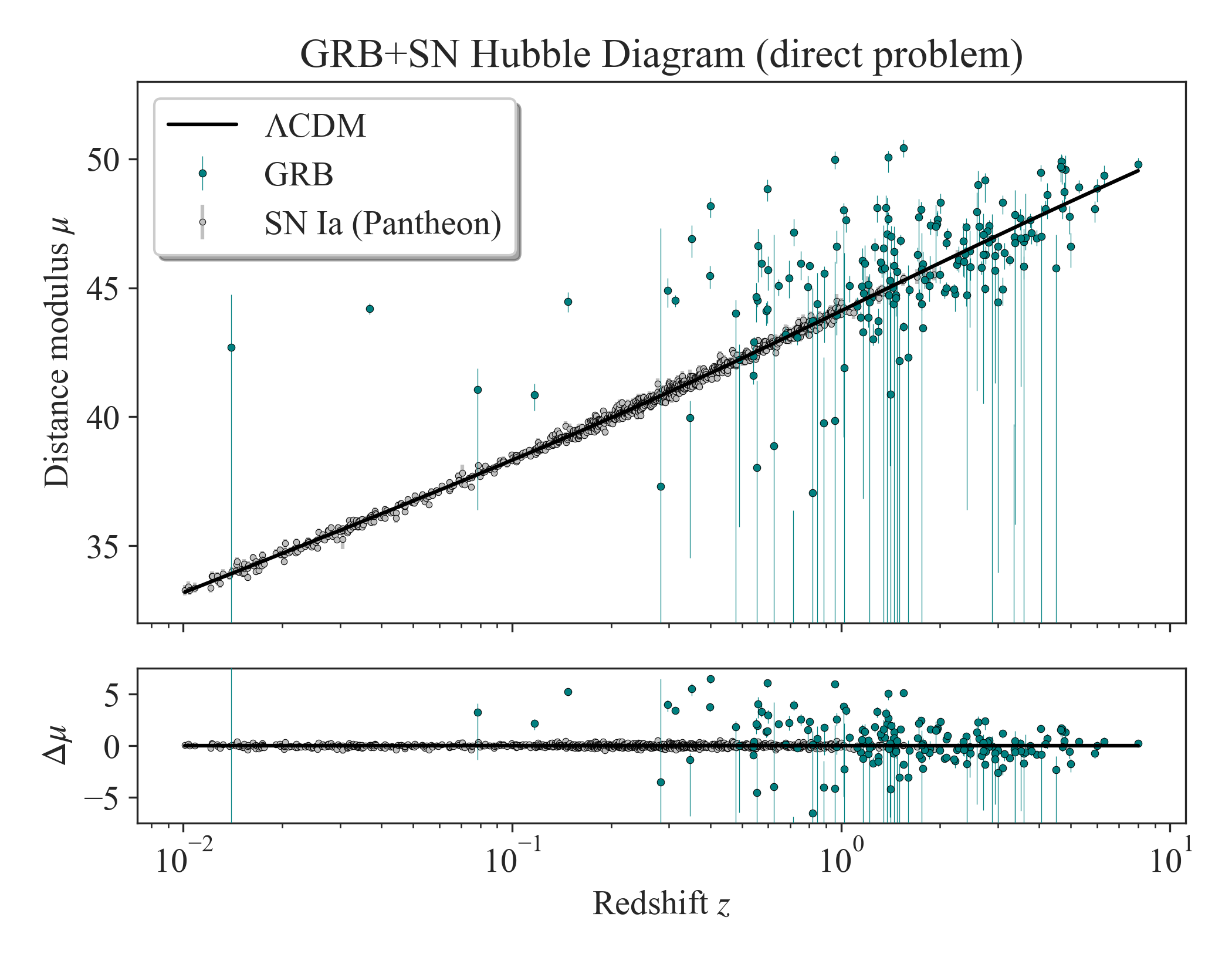}
    % \label{fig:HDts}
    \caption{The results of the direct problem algorithm with fixed $k=0$. 
    Top: 
    the Monte Carlo distributions for the parameters $a$ and $b$, obtained as a result of linear approximation by the Theil-Sen method in the $\log \Eiso$ -- $\log \Epi$ plane. The estimates of these parameters with their errors are shown.
    Middle: the $\log \Eiso$ -- $\log \Epi$ plane.
    Bottom: the joint SN Ia and LGRB HD.} 
    \label{fig:direct}
\end{figure}

To solve the direct problem, we used the previously obtained $\Lambda$CDM parameters $\Omm=0.296$, $\OmL=0.704$, and $k=0$. In this context, the Amati parameters $a$ and $b$ were determined through linear approximation employing the Theil-Sen method. More precisely, by using our Monte-Carlo sample with a size of $N=10,000$ (as detailed in Section~\ref{subsec:montecarlo}), we acquired $10,000$ sets of $a$ and $b$ values using the Theil-Sen method. From these sets, we calculated the medians and quantiles for the Amati parameters: $a=0.92^{+0.10}_{-0.11}$ and $b=50.48^{+0.28}_{-0.27}$. With these values, we constructed the joint SN Ia and LGRB HD. The results are shown in Figure~\ref{fig:direct}. The algorithm provides optimal values for the Amati parameters and extends the HD at the LGRB scales.

\subsection{Inverse cosmological calibration problem}
\label{subsec:inverse}

\begin{figure*}
    \centering
    \includegraphics[width=0.49\textwidth]{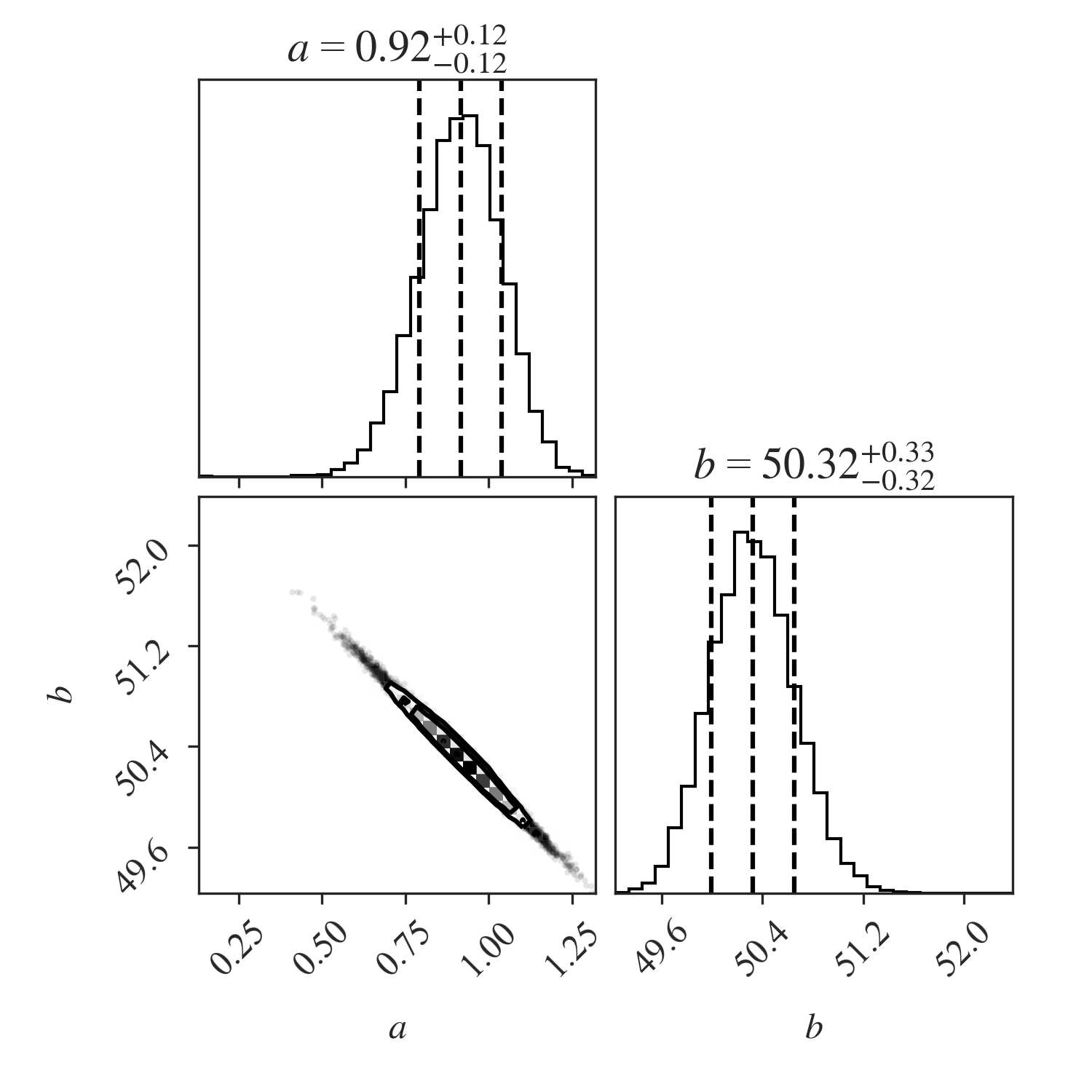}
        % \label{fig:cornericf0}
    \includegraphics[width=0.49\textwidth]{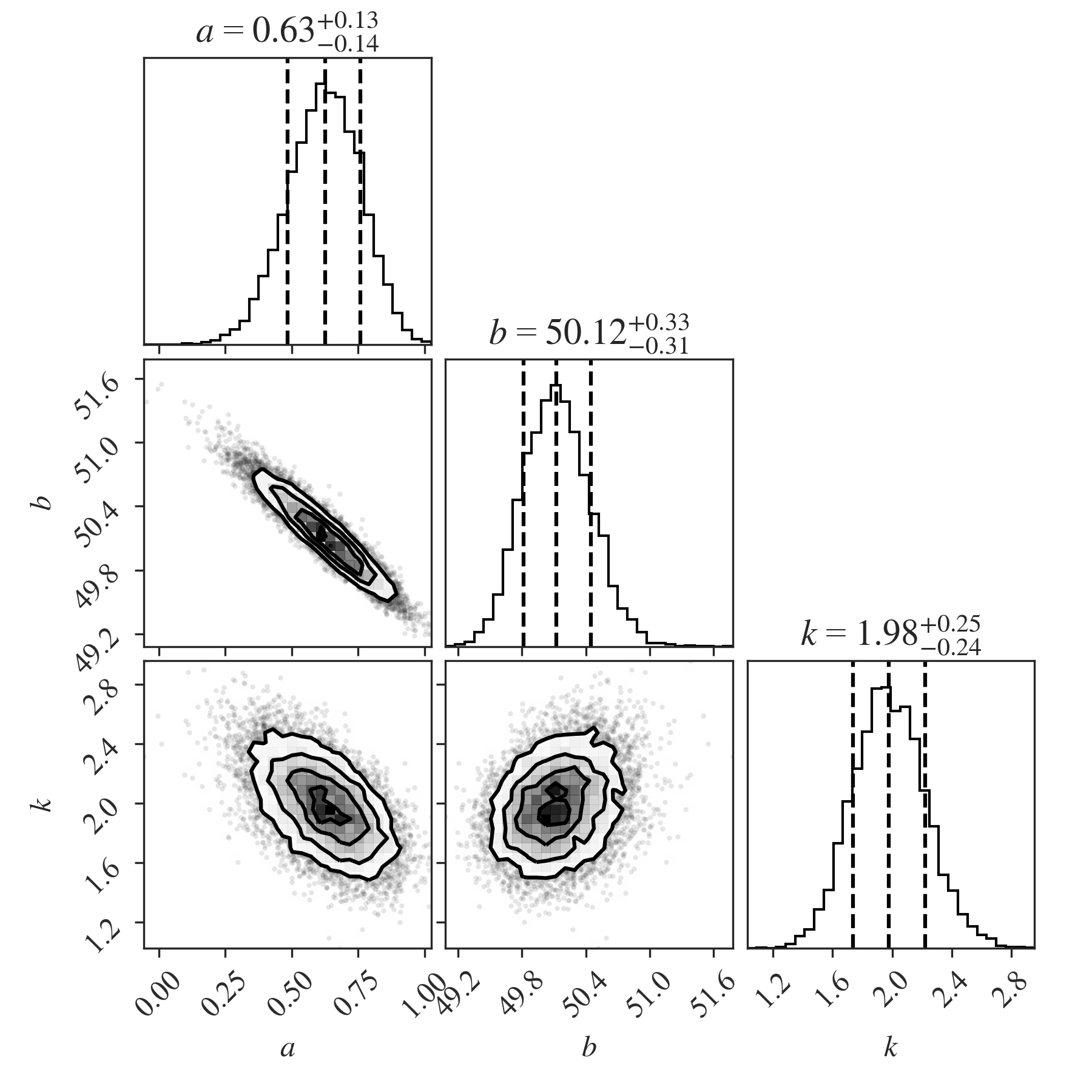}
        % \label{fig:cornericf}
    \includegraphics[width=0.48\textwidth]{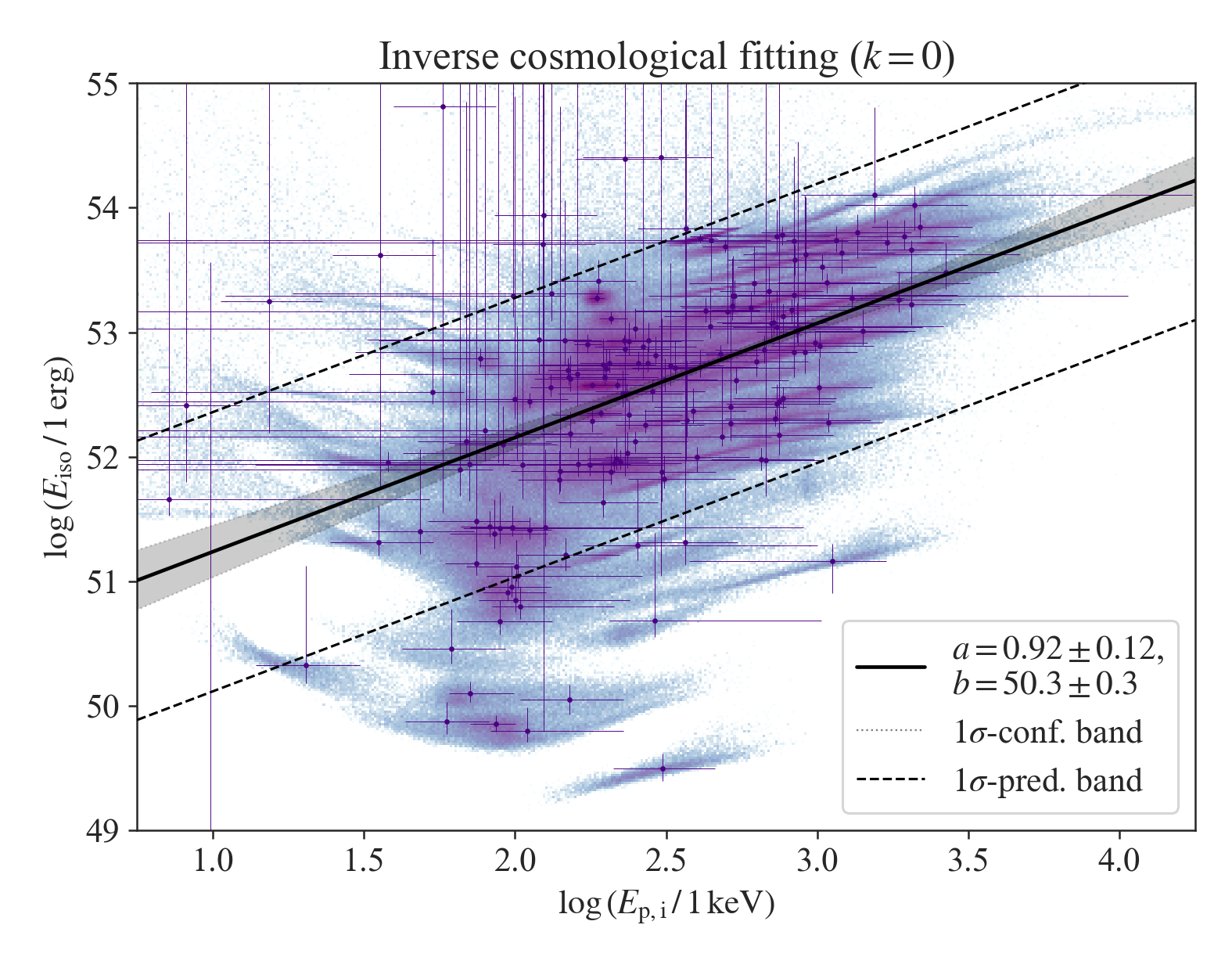}
        % \label{fig:Amatiicf0}
    \includegraphics[width=0.48\textwidth]{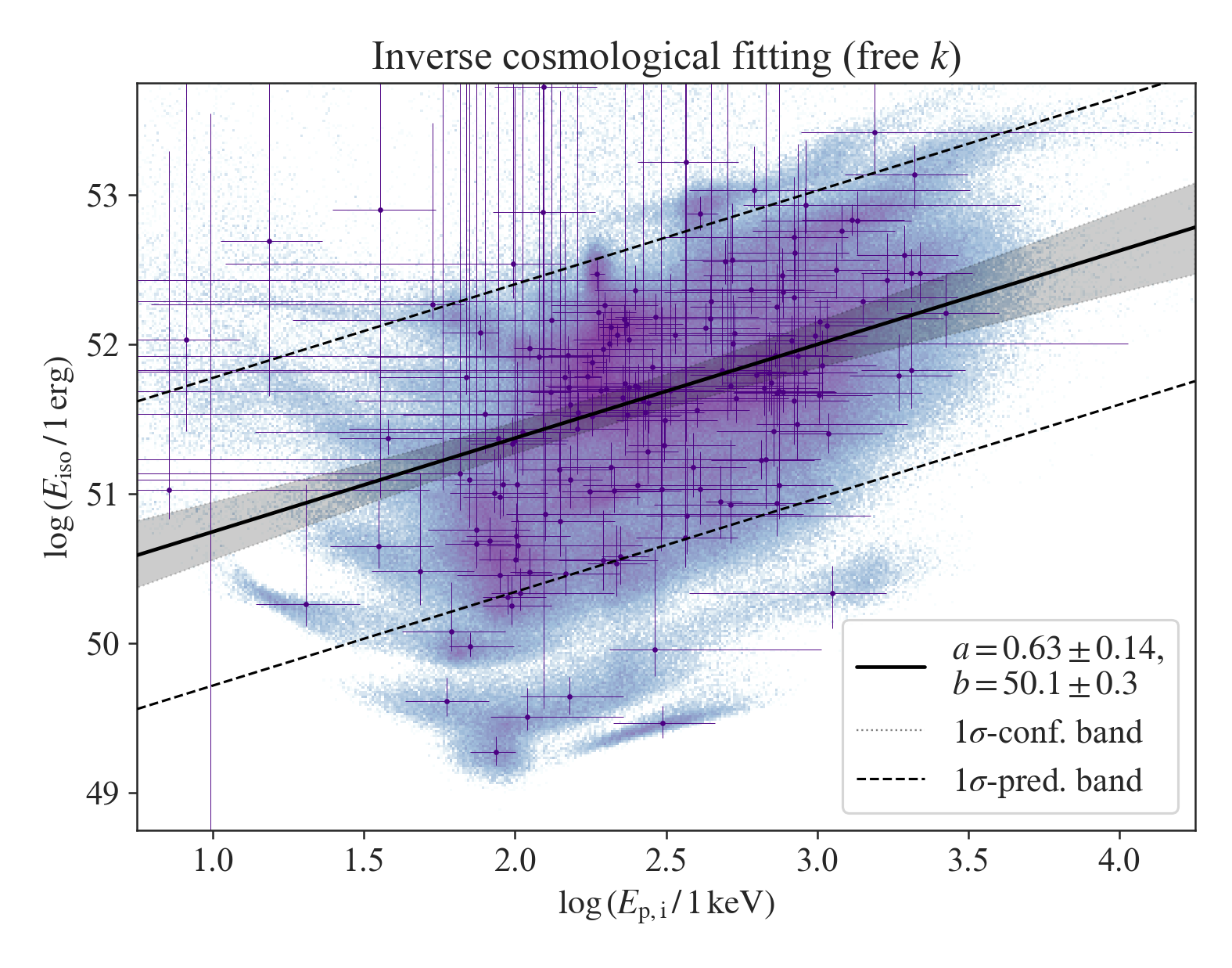}
        % \label{fig:Amatiicf}
    \includegraphics[width=0.48\textwidth]{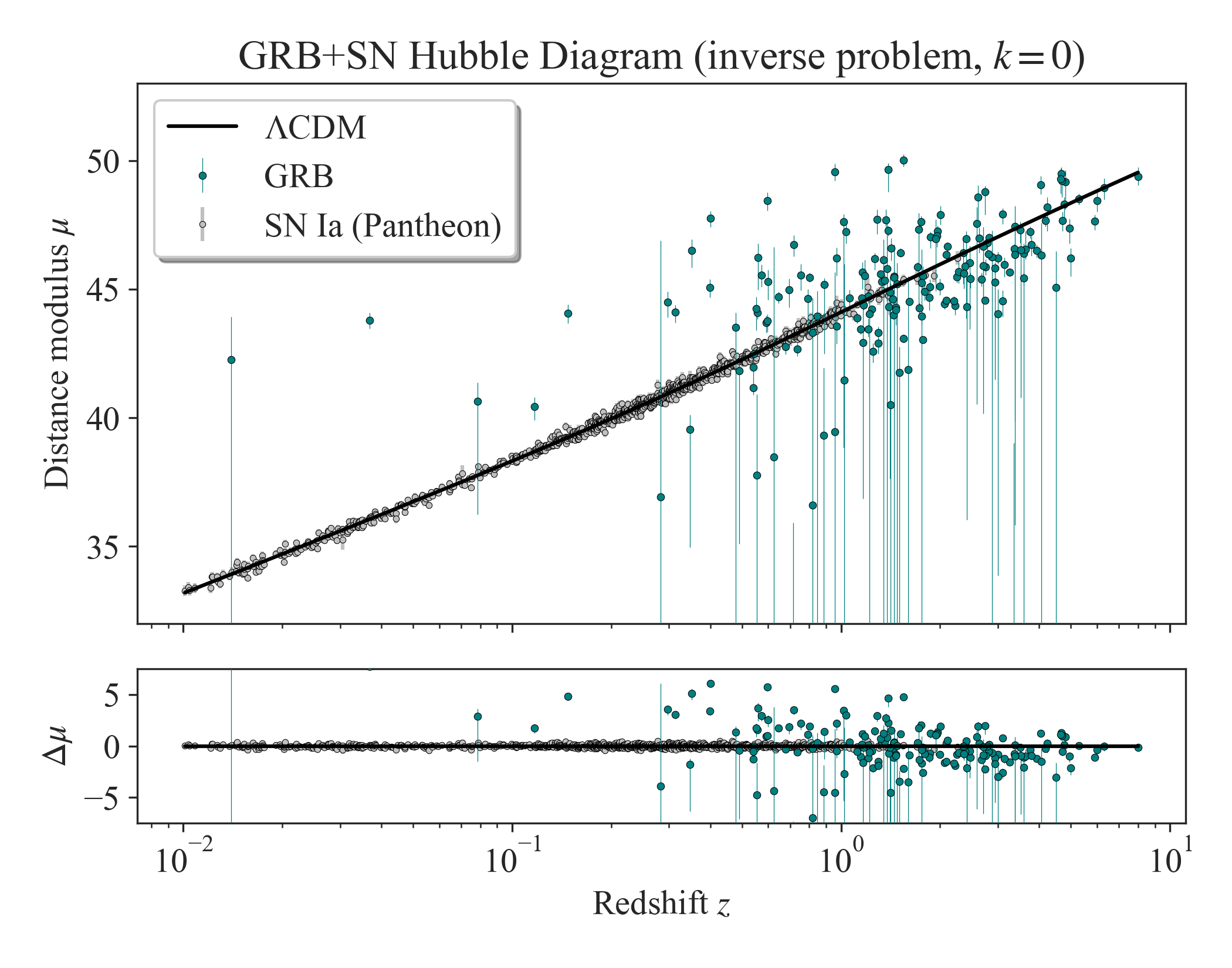}
        % \label{fig:HDicf0}
    \includegraphics[width=0.48\textwidth]{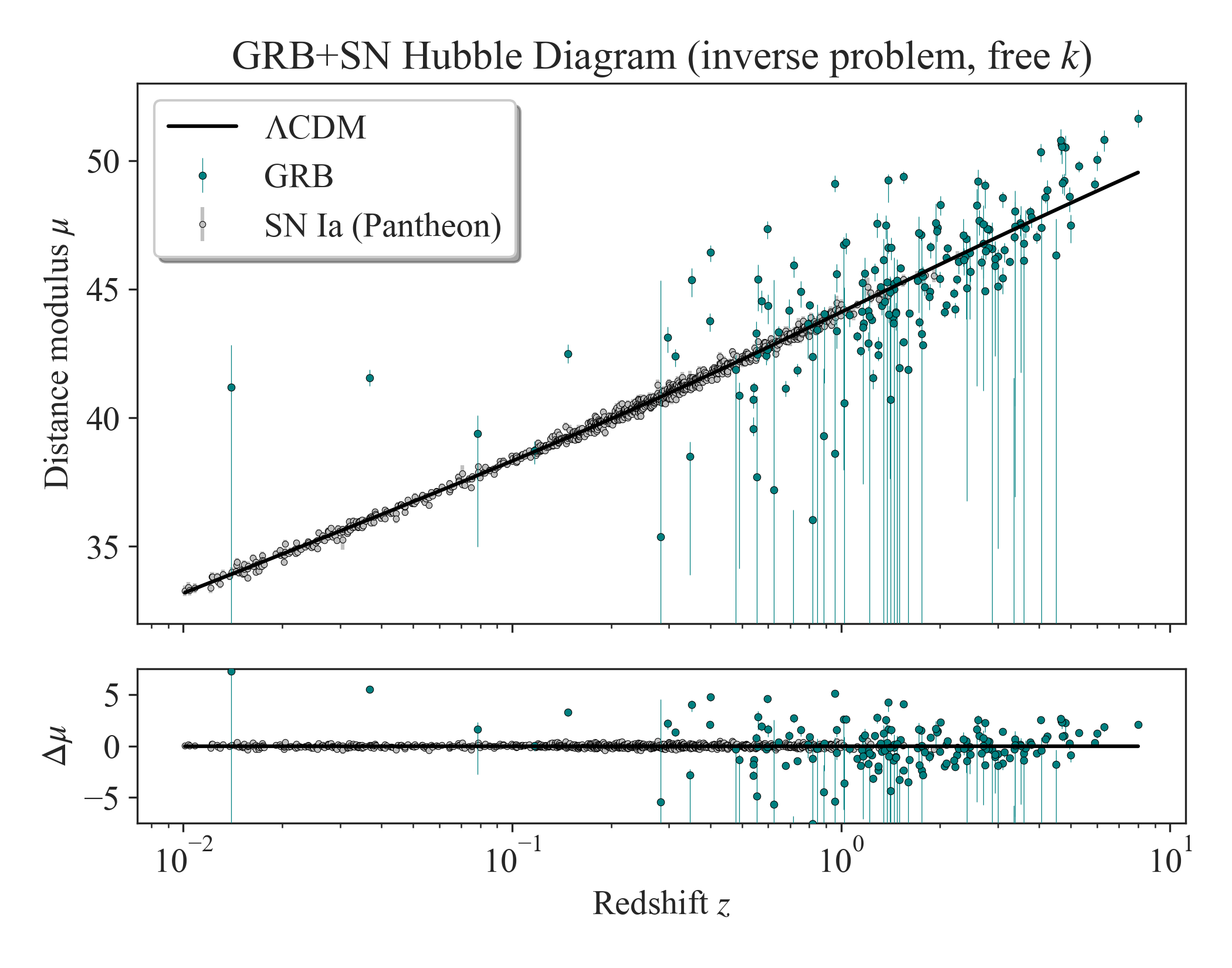}
        % \label{fig:HDicf}
    \caption{
    The results of the ICCP algorithm. Left: the case of fixed $k=0$. Right: all three parameters $a, b, k$ are varied.
    Top: The Monte Carlo distributions for the varied parameters, obtained as a result of solving the ICCP (minimising the deviations of $\muagrb$ (Equation~(\ref{eq:Amatiext})) from the $\Lambda$CDM model $\mu(z)$ (Equation~(\ref{eq:mu})) on the HD). 
    Middle: the $\log \Eiso$ -- $\log \Epi$ plane.
    Bottom: the joint SN Ia and LGRB HD.}
    \label{fig:inverse}
\end{figure*}

The results of our algorithm for solving the ICCP case (as detailed in Section \ref{sec:ICCP}) are shown in Figure~\ref{fig:inverse}. We have regarded two models: one with a fixed $k=0$ and the other with a free $k$. The input cosmological model is based on the $\Lambda$CDM parameters obtained in Section \ref{sec:snlcdm}. 
In the ICCP case with $k=0$, the results coincide within the margin of errors with the results obtained for the direct problem with the same value of $k=0$.
Thus, considering the HD to minimize residuals has minimal impact on the Amati parameter values. However, when we introduce variation in the parameter $k$, its influence becomes evident, with an optimal value of $k=2$ and a reduction in the correlation in the $\log \Eiso$ -- $\log \Epi$ plane. Currently, the Monte-Carlo samples has a more regular configuration, and the joint SN Ia and GRB HD appears visually more regular as well.

\section{Conclusions}

The sample of 174 LGRBs from the Swift mission was calibrated using the non-parametric statistical methods (partially described in~\citet{Lovyagin2022} and \citet{gorokhov2023nonparametric}), the $\Lambda$CDM model as potentially changeable basis, and the extended Amati relation with cosmological correction for gravitational lensing and Malmquist bias (GLMB) from \citet{shirokov2020high}.
The calibration of the Amati parameters and the GLMB correction ($a,\,b,\,k$) was carried out in such a way that residuals in the $\mu$ -- $z$ diagram (the HD) were minimised, while its points $\mu_i$ were calculated from the $\log \Epi$ -- $\log \Eiso$ plot (the Amati plane) that was varied by ($a,\,b,\,k$).
This way we called as inverse cosmological calibration problem (ICCP). Thus, we suggest the ICCP as a new method resolving the circularity problem of the LGRB calibration as standard candles.  In this research we used the $\Lambda$CDM model parameters found from the type Ia supernovae HD from the Pantheon catalogue, except for the Hubble constant which was fixed at $H_0=70$ km/s/Mpc. 

Two cases were tested, one with a fixed GLMB parameter $k=0$ (the case of no GLMB correction) and the second with variation of all three parameters $a,\, b,\, k$ that is of interest to us. Additionally, we have compared the case of ICCP solution with $k=0$ to the solution of the direct cosmological problem (we called an estimation of $a$ and $b$ using linear regression in the $\log \Eiso$ -- $\log \Epi$ plane with a fixed cosmological model as a direct problem). Although the illustrative solution of the direct problem suffers from the circularity problem, it can be seen that the results are very similar, the parameters coincide up to an error. Thus, we can conclude that our methods and determination of the parameters of the $\Lambda$CDM model were accurate (including the fact that parameters of $\Lambda$CDM model fit the HD for LGRB under the conditions of this problem within the current limits of accuracy).  

Adding a third degree of freedom in the form of a correction parameter $k$ affecting the bolometric fluence, given by Equation (\ref{eq:correction}), and hence the extended Amati relation (\ref{eq:Amatiext}), causes a significant upward bias on the factor $(z+1)^2$ in estimates of the LGRB distance moduli $\mu_i$ in the HD.
While the Amati parameters $a=0.92^{+0.12}_{-0.12}$, $b=50.32^{+0.33}_{-0.32}$ for $k=0$ are changing in the direction of decreasing the correlation with $a=0.63^{+0.13}_{-0.14}$, $b=50.12^{+0.33}_{-0.31}$ for $k=1.98^{+0.25}_{-0.24}$.
All estimates of the Amati parameters are presented in Table~\ref{tab:finalres}. 
It is natural to assume that the Amati correlation of the bolometric fluence ($\Eiso$) with its hardness ($\Epi$) depends on the mass of the collapsing star. Therefore, for physical reasons, the correlation should be related to the evolution of stars, and also serve as a test of alternative ideas, such as the existence of quark stars~\citep{Sokolov2015,Sokolov2016,Sokolov2019} and test of gravitational theories~\citep{baryshev2020einstein,shirokov2020theseus}.

% Таблицу и значения в тексте исправить после пересчета
% ======
\begin{table}
    \centering
    \begin{tabular}{r|ccc}
    \hline
        The Theil-Sen method & $a$ & $b$ & $k$ \\ \hline
        Direct problem, $k=0$ & $0.92\pm0.10$ & $50.5 \pm 0.3$ & $0$ \\
        Inverse problem, $k=0$ & $0.92\pm0.12$ & $50.3\pm0.3$ & $0$ \\
        Inverse problem, free $k$ & $0.63\pm0.13$ & $50.1\pm0.3$ & $2.0\pm0.2$ \\ \hline
     \end{tabular}
     \caption{Estimation of the Amati parameters $a, b$ and the GLMB correction parameter $k$.}
    \label{tab:finalres}
\end{table}
% =======

The accuracy of the observational values of LGRBs, as well as the number of their observations at the moment, leave much to be desired. 
However, we conclude that LGRBs can be used as standard candles, allowing one to expand the HD up to $z\approx10$, and their position on the HD can be correctly consistent with the selected cosmological model.
Apparently, the standard $\Lambda$CDM model is indeed consistent with the indirect estimates of the LGRB HD, however, this requires further research using our method and varied models as input basis, as well as increasing the sample size of LGRBs. 
As an application of our methods for research we suggest addressing the circularity problem analogically \citet{amati2013, demianski2021prospects} via iterations and solving the inverse problem to estimate parameters of cosmological models basing on the residuals values of observational data in the Bayesian statistics paradigm. Thus, the ICCP idea can be used as an alternative cosmological model test in nearest future. 

\section*{Acknowledgements}
% The Acknowledgements section is not numbered. Here you can thank helpful colleagues, acknowledge funding agencies, telescopes and facilities used etc. Try to keep it short.

We thank the anonymous reviewer for important suggestions that helped us to improve the presentation of our results.
% We are grateful to  for useful discussions and comments.
Part of the observational data was exposured on the unique scientific facility the Big Telescope Alt-azimuthal SAO RAS and the data processing was supported under the Ministry of Science and Higher Education of the Russian Federation grant 075-15-2022-262 (13.MNPMU.21.0003).

%%%%%%%%%%%%%%%%%%%%%%%%%%%%%%%%%%%%%%%%%%%%%%%%%%
\section*{Data Availability}
% The inclusion of a Data Availability Statement is a requirement for articles published in MNRAS. Data Availability Statements provide a standardised format for readers to understand the availability of data underlying the research results described in the article. The statement may refer to original data generated in the course of the study or to third-party data analysed in the article. The statement should describe and provide means of access, where possible, by linking to the data or providing the required accession numbers for the relevant databases or DOIs.

The codes developed in Python underlying this article are available in the repository on \url{https://github.com/Roustique/sngrb}.



\bibliographystyle{mnras}
\bibliography{bibliography}

%\begin{thebibliography}{99}
%\end{thebibliography}

%%%%%%%%%%%%%%%%%%%%%%%%%%%%%%%%%%%%%%%%%%%%%%%%%%

%%%%%%%%%%%%%%%%% APPENDICES %%%%%%%%%%%%%%%%%%%%%

\appendix

\section{abbreviations}
\label{app:A}

\itemize{

\item HD      -- Hubble diagram                
    % introduced on  line
\item SN(e)   -- supernova(e)    
    % introduced on  line
\item GRB(s)  -- gamma-ray burst(s)     
    % introduced on  line
\item LGRB(s) -- long gamma-ray Burst(s)
    % introduced on  line
% \item FLRW      -- Friedmann--Lemaitre--Robertson--Walker
    % introduced on  line
\item SCM     -- standard cosmological model
    % introduced on  line
% \item THESEUS -- Transient High-Energy Sky and Early Universe Surveyor (space mission)
    % introduced on  line
% \item Euclid  -- space mission
    % introduced on  line
\item GLMB    -- gravitational lensing and Malmquist bias
    % introduced on  line
% \item MB  --   Malmquist bias      
    % introduced on  line
% \item GL        -- gravitational lensing
    % introduced on  line
% \item GLB    --   gravitational lensing bias
    % introduced on  line
% \item GRT       -- general relativity theory
    % introduced on  line
% \item ECP  --  Einstein's cosmological principle
    % introduced on  line
% \item EoS(s)   -- Equation of state(s)
    % introduced on  line
% \item QSO(s)   -- quasi-stellar object(s)
    % introduced on  line
%\item PDF       -- Probability Distribution Function
    % introduced on  line
}

\section{Original LGRB catalogue}
\label{app:B}
% \addcontentsline{toc}{subsection}{A. Original catalogue of LGRBs}

\onecolumn

\begin{center}
\footnotesize
\begin{longtable}{c||ccc|ccc|cc|c}
\hline
    LGRB & $\alpha$ & $\Delta_{\mathrm{l}}\alpha$ & $\Delta_{\mathrm{u}}\alpha$ & $\Ep$ & $\Delta_{\mathrm{l}}\Ep$ & $\Delta_{\mathrm{u}}\Ep$ & $\Sobs$ & $\Delta_{90\%}\Sobs$ & $z$ \\
\hline
111209A &     -1.447 &      0.05384 &       0.1206 &   768.8 &        &        &               360 &             10 &     0.677 \\
150314A &    -0.9983 &      0.04947 &      0.09653 &   764.7 &     320.6 &        &    220 &   3 &     1.758 \\
200829A &    -0.9194 &      0.05062 &       0.1023 &   694.2 &     303.8 &      7233 &    510 &      10 &      1.25 \\
170604A &     -1.224 &       0.1804 &       0.3939 &   607.4 &        &        &     51 &              4 &     1.329 \\
191004B &     -1.007 &       0.2229 &       0.6018 &   593.6 &        &        &     25 &              1 &     3.503 \\
141109A &     -1.456 &       0.0849 &       0.2461 &     512 &        &        &     68 &              3 &     2.993 \\
181020A &     -1.159 &       0.1206 &       0.2381 &   495.4 &        &        &     81 &              3 &     2.938 \\
\hline
\end{longtable}
\end{center}

\section{Enhanced LGRB catalogue}
\label{app:C}
% \addcontentsline{toc}{subsection}{B. Enhanced catalogue of LGRBs}

\begin{center}
\footnotesize
\begin{longtable}{c||ccc|ccc|cc|c||ccc}
\hline
    LGRB & $\alpha$ & $\Delta_{\mathrm{l}}\alpha$ & $\Delta_{\mathrm{u}}\alpha$ & $\Ep$ & $\Delta_{\mathrm{l}}\Ep$ & $\Delta_{\mathrm{u}}\Ep$ & $\Sobs$ & $\Delta\Sobs$ & $z$ & $\muagrb$(direct) & $\muagrb(k=0)$ & $\muagrb$(free $k$) \\
\hline
111209A &     -1.447 &      0.05384 &       0.1206 &   768.8 &     239.4 &     385.2 &    360 &           6.08 &     0.677  & $43.18^{+0.288}_{-0.242}$ & $42.77^{+0.273}_{-0.218}$ & $41.13^{+0.313}_{-0.305}$ \\
150314A &    -0.9983 &      0.04947 &      0.09653 &   764.7 &     320.6 &     383.1 &    220 &          1.824 &     1.758  & $44.37^{+0.247}_{-0.235}$ & $43.95^{+0.237}_{-0.211}$ & $43.27^{+0.244}_{-0.314}$ \\
200829A &    -0.9194 &      0.05062 &       0.1023 &   694.2 &     303.8 &      7233 &    510 &           6.08 &      1.25  & $43.01^{+0.586}_{-0.237}$ & $42.59^{+0.645}_{-0.204}$ & $41.55^{+0.32}_{-0.43}$ \\
170604A &     -1.224 &       0.1804 &       0.3939 &   607.4 &     189.1 &     304.3 &     51 &          2.432 &     1.329  & $45.71^{+0.301}_{-0.31}$ & $45.3^{+0.282}_{-0.288}$ & $44.35^{+0.276}_{-0.348}$ \\
191004B &     -1.007 &       0.2229 &       0.6018 &   593.6 &     184.8 &     297.4 &     25 &          0.608 &     3.503  & $47.7^{+0.369}_{-0.468}$ & $47.29^{+0.359}_{-0.452}$ & $47.57^{+0.319}_{-0.49}$ \\
141109A &     -1.456 &       0.0849 &       0.2461 &     512 &     159.4 &     256.5 &     68 &          1.824 &     2.993  & $46.61^{+0.347}_{-0.295}$ & $46.2^{+0.34}_{-0.286}$ & $46.28^{+0.258}_{-0.229}$ \\
181020A &     -1.159 &       0.1206 &       0.2381 &   495.4 &     154.2 &     248.2 &     81 &          1.824 &     2.938  & $46.25^{+0.289}_{-0.257}$ & $45.83^{+0.282}_{-0.239}$ & $45.91^{+0.22}_{-0.243}$ \\
\hline
\end{longtable}
\end{center}

\twocolumn

% Don't change these lines
\bsp	% typesetting comment
\label{lastpage}
\end{document}